\documentclass{emulateapj}

\usepackage{multirow}
\usepackage{hyperref}

\newcommand{\beq}{\begin{equation}}
\newcommand{\eeq}{\end{equation}}

\def\earth{\hbox{$\oplus$}}
\def\arcmin{\hbox{$^\prime$}}

\def\deg{\hbox{$^\circ$}}
\newcommand{\lsim}{\ \raise
-2.truept\hbox{\rlap{\hbox{$\sim$}}\raise5.truept\hbox{$<$}\ }}
\newcommand{\gsim}{\ \raise
-2.truept\hbox{\rlap{\hbox{$\sim$}}\raise5.truept\hbox{$>$}\ }}
\newcommand{\simsim}{\ \raise
-2.truept\hbox{\rlap{\hbox{$\sim$}}\raise5.truept\hbox{$\sim$}\ }}

\renewcommand{\arraystretch}{1.2}

\slugcomment{ApJ accepted}
\shorttitle{IN-SYNC Orion}
\shortauthors{Da Rio et al. 2017}

\begin{document}

\title{IN-SYNC. V. Stellar kinematics and dynamics in the Orion A Molecular Cloud}

\author{
Nicola~Da Rio$^1$,
Jonathan~C.~Tan$^{1,2}$,
Kevin~R.~Covey$^3$,
Michiel~Cottaar$^4$,
Jonathan~B.~Foster$^5$,
Nicholas~C.~Cullen$^1$,
John~Tobin$^{6,14}$,
Jinyoung~S.~Kim$^7$,
Michael~R.~Meyer$^8$,
David~L.~Nidever$^8$,
Keivan~G.~Stassun$^{9}$,
S.~Drew~Chojnowski$^{10}$,
Kevin~M.~Flaherty$^{11}$,
Steven~R.~Majewski$^{10}$,
Michael~F.~Skrutskie$^{10}$,
Gail~Zasowski$^{10,12,13}$,
Kaike~Pan$^{15}$}

\affil{ \\
$^1$Department of Astronomy, University of Florida, Gainesville, FL 32611, USA. \\
$^2$Department of Physics, University of Florida, Gainesville, FL 32611, USA. \\
$^3$Department of Physics \& Astronomy, Western Washington University, Bellingham, WA 98225, USA. \\
$^4$Department of Clinical Neurosciences, University of Oxford, Oxford, United Kingdom. \\
$^5$Yale Center for Astronomy and Astrophysics, Yale University New Haven, CT 06520, USA. \\
$^6$Leiden Observatory, NL-2333CA Leiden, The Netherlands \\
$^7$Steward Observatory, University of Arizona, Tucson, AZ 85721, USA. \\
$^8$Department of Astronomy, University of Michigan, Ann Arbor, MI 48109, USA.\\
$^{9}$Department of Physics \& Astronomy, Vanderbilt University, Nashville, TN 37235, USA.\\
$^{10}$Department of Astronomy, University of Virginia, Charlottesville, VA 22904, USA. \\
$^{11}$Astronomy Department, Wesleyan University, Middletown, CT 06459, USA.\\
$^{12}$Department of Astronomy, The Ohio State University, Columbus, OH 43210, USA.\\
$^{13}$Center for Cosmology and Astro-Particle Physics, The Ohio State University, Columbus, OH 43210, USA.\\
$^{14}$Homer L. Dodge Department of Physics and Astronomy, University of Oklahoma, 440 W. Brooks Street, Norman, OK 73019, USA. \\
$^{15}$Apache Point Observatory and New Mexico State University, P.O. Box 59, Sunspot, NM 88349-0059, USA.}

\email{ndario@ufl.edu}


\begin{abstract}
The kinematics and dynamics of young stellar populations enable us to test theories of star formation. With this aim, we continue our analysis of the SDSS-III/APOGEE IN-SYNC survey, a high resolution near infrared spectroscopic survey of young clusters. We focus on the Orion A star-forming region, for which IN-SYNC obtained spectra of $\sim2700$ stars. In Paper IV we used these data to study the young stellar population. Here we study the kinematic properties through radial velocities ($v_r$). The young stellar population remains kinematically associated with the molecular gas, following a $\sim10\:{\rm{km\:s}}^{-1}$ gradient along filament. However, near the center of the region, the $v_r$ distribution is slightly blueshifted and asymmetric; we suggest that this population, which is older, is slightly in foreground. We find evidence for kinematic subclustering, detecting statistically significant groupings of co-located stars with coherent motions. These are mostly in the lower-density regions of the cloud, while the ONC radial velocities are smoothly distributed, consistent with it being an older, more dynamically evolved cluster. The velocity dispersion $\sigma_v$ varies along the filament. The ONC appears virialized, or just slightly supervirial, consistent with an old dynamical age. Here there is also some evidence for on-going expansion, from a $v_r$--extinction correlation. In the southern filament, $\sigma_v$ is $\sim2$--$3$ times larger than virial in the L1641N region, where we infer a superposition along the line of sight of stellar sub-populations, detached from the gas. On the contrary, $\sigma_v$ decreases towards L1641S, where the population is again in agreement with a virial state.
\end{abstract}

\keywords{stars: formation, pre-main sequence, kinematics and dynamics; open clusters and associations: individual (Orion Nebula Cluster, L1641)  }


\section{Introduction}
\label{section:introduction}

The observational study of stellar kinematics in young clusters
provides critical clues on the mechanisms governing cluster formation
and early evolution. The velocities of young stars, in comparison with
gas kinematics, can reveal if and for how long the newly formed
stellar population follows the initial/early gas flows, which may be
due to turbulence, gravitational infall, cloud collisions or some
other triggering mechanism. The star cluster may be forming in a quasi
monolithic fashion or the process may involve merging of different
sub-clusters. These mechanisms may be revealed by a study of the
kinematics, including kinematic substructure, of the young
stars. Subsequent dynamical evolution may involve mixing of orbits,
e.g., as the star cluster virializes as a gravitationally bound
system. Alternatively, depending on the overall efficiency of
formation from the natal gas clump and degree of gas expulsion, the
stellar population may be or become unbound and thus expand with less
opportunity for stellar interactions. Depending on the rate of star
formation, i.e., the efficiency per local free-fall time, these
potential evolutionary pathways may be followed even while the
cluster/clump system is still gas dominated.  Another way to test
different star formation scenarios is thus to measure the velocity
dispersion of the cluster and compare it with the value expected from
virial equilibrium given the gravitational potential due to the total
mass (stars + gas) of the system.

\begin{figure*}
\epsscale{1.1}
\plotone{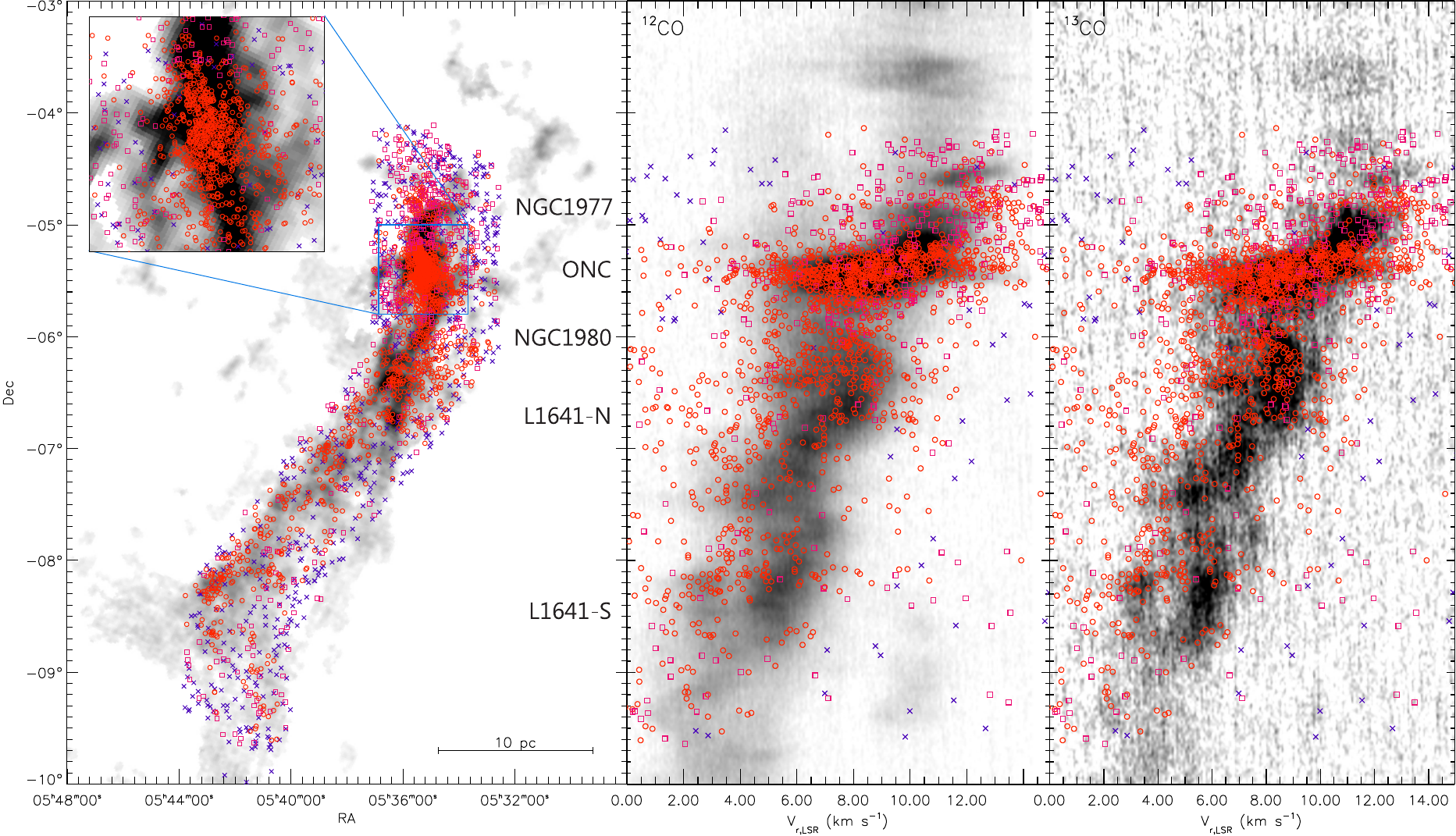}
\caption{{\em Left panel:} spatial distribution of the IN-SYNC targets, overplotted on a $^{13}$CO(2-1) map from \citet{nishimura2015}. Red circles indicate known members from the literature, magenta squares new candidate members from Paper IV, blue crosses remaining sources, likely non-members. {\em Middle and right panels}: position-velocity diagram for the targets, compared to either $^{12}$CO(2-1) or $^{13}$CO(2-1) data. \label{figure:radec}}
\end{figure*}


\begin{figure*}
\epsscale{1.1}
\includegraphics[width=\textwidth]{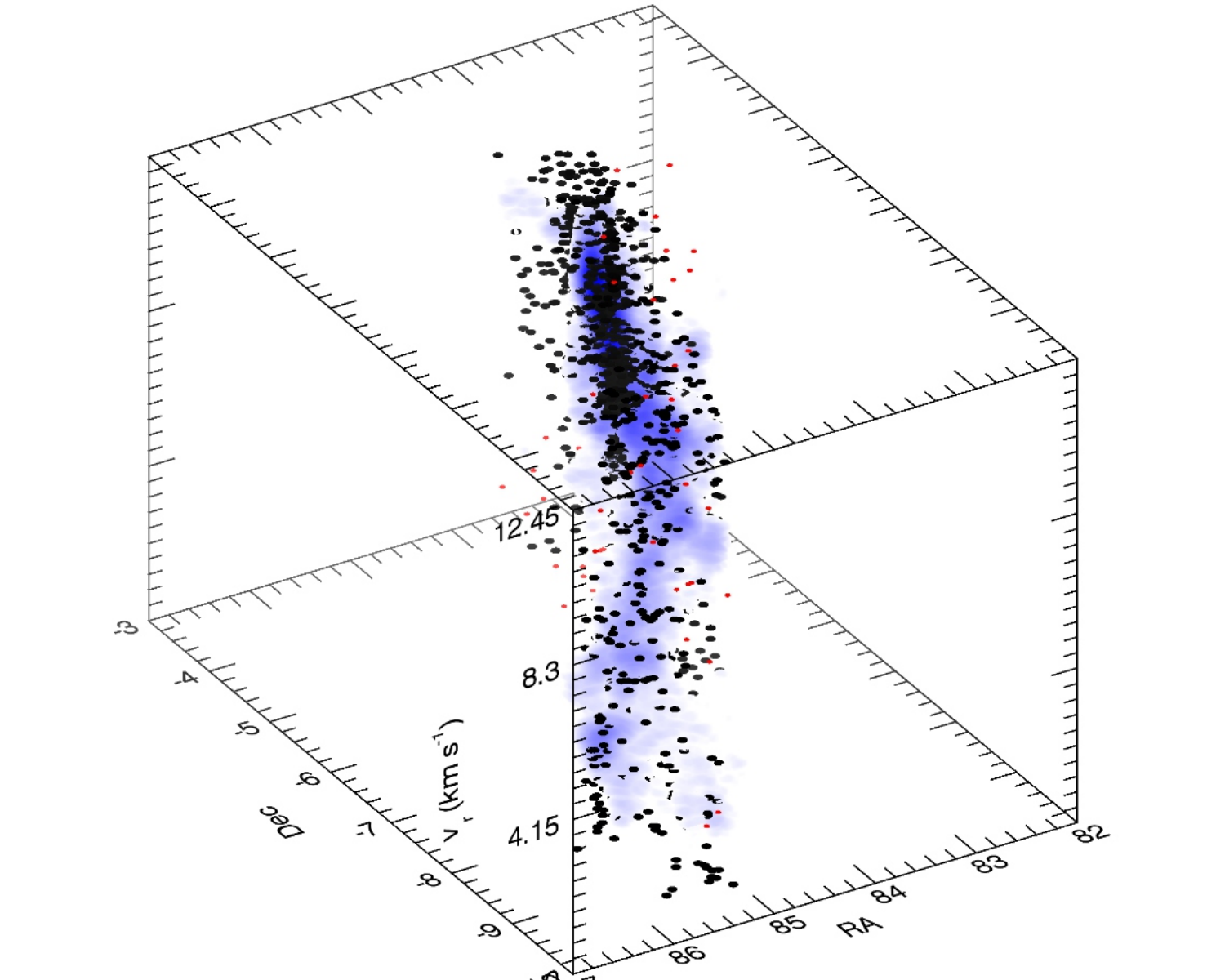}
\caption{Animation showing the 3-dimensional position-position-velocity diagram of our stellar sample, compared to that of the $^{13}$CO.
Full animation available \href{https://www.dropbox.com/s/o7yjrqpo87kckam/figure2_video.avi?dl=0}{here}.\label{figure:3dvideo}}
\end{figure*}

Great effort has been made over the years to study the dynamical
evolution of young clusters theoretically, through N-body simulations
\citep[e.g.,][]{scally2002,goodwin2006,baumgardt2007,fellhauer2009,allison2009,allison2010,parker2012,parker2014,farias2015,farias2017}. Results
reveal a fairly complex picture, in that the dynamical state and
evolution of the stellar population, the bound fraction and morphology
depend on several factors: the star formation efficiency, the time of
gas removal and its duration and the initial configuration of
hierarchical/substructured protoclusters.

Observationally, some works have analyzed the morphology of young
clusters at different ages to constrain their dynamical evolution and
initial conditions
\citep[e.g.,][]{gutermuth2005,schmeja2006,banerjee2015,dario2014b,jaehnig2015}.
However, observations of radial velocities in young clusters are a
more powerful tool to probe the current dynamical state, search for
and parameterize binary populations, and investigate spatially
coherent velocity gradients or substructure that might give clues to
the initial conditions or reveal multiple populations
\citep{furesz2008,tobin2009,cottaar2012,jeffries2006,jeffries2014,rigliaco2016}. {In many cases, these
radial velocities, $v_r$, have been obtained from optical
high-resolution spectroscopy surveys that reach radial velocity precisions of
$\sim1$~km~s$^{-1}$ or better for individual sources. Such precision is
needed to resolve the kinematics of nearby,
relatively low-mass regions of star formation (e.g., Orion), which
have velocity dispersions of a few km~s$^{-1}$. In young embedded clusters, optical spectroscopy
can be limited by dust extinction, thus unable to reach stellar members affected by high extinction. }

In this context, with the INfrared Spectra of Young Nebulous Clusters
(IN-SYNC) project \citet{cottaar2014}, an ancillary program of the
SDSS-III, we obtained multiobject, near-infrared (NIR) high-resolution
spectra in 3 young clusters. This program, which used Apache Point
Observatory Galactic Evolution Experiment (APOGEE) spectrograph,
allowed us to derive stellar parameters and reaching radial velocity
precisions to $\sim0.3$~km~s$^{-1}$ down to $H=12.5$~mag. The first
part of this survey covered the Perseus cloud, through their young
clusters IC348 \citep[][hereafter Paper~I and
  Paper~III]{cottaar2014,cottaar2015} and NGC~1333 \citep[][hereafter
  Paper~II]{foster2015}. We have found evidence for a supervirial
stellar population in IC 348, whereas in NGC 1333 the stars are in
close agreement with virial velocities, but the diffuse gas and dense
gas have significantly different velocity dispersion. Moreover, we
found that stars in NGC 1333 show a similar velocity dispersion to the
diffuse gas, whereas the dense cores appear in subvirial motions,
unless their dynamics are significantly regulated by large scale
magnetic fields.

In a second part of IN-SYNC, we covered the Orion A molecular cloud. At the distance of $\sim400$~pc  \citep{menten2007}, this filamentary structure includes the Orion Nebula Cluster (ONC), the closest site of active massive star formation. Sparser populations are present to the north, in the upper sword and NGC 1977, and to the south, with NGC~1980 and L1641. This population has been thoroughly studied at many wavelengths through photometry and spectroscopy \citep{hillenbrand1997,hillenbrand-hartmann1998,dario2010a,dario2012,robberto2010,robberto2013,hsu2012,hsu2013,fang2009,fang2013}. Memberships have been estimated through optical spectroscopy, presence of IR excess emission \citep{megeath2012}, X-ray emission \citep{getman2005,pillitteri2013}. The population has a typical model dependent mean age of $\sim2-3$~Myr throughout the region, and in spite of the large extent of the structure ($\sim40$~pc). The only exception is the population around NGC1980 being somewhat older \citep{alves2012,bouy2014,dario2016}. In the ONC, a large age spread compared to the local free-fall time is present \citet{dario2010a,dario2014a,dario2016}, hinting to a slow star formation process relative to the dynamical timescale of this star-forming clouds \citep{tan2006,krumholz2007}.

The kinematic properties of the molecular gas in Orion A have been
studied in great detail
\citep[e.g.,][]{bally1987,dame2001,nishimura2015,ishii2016} from
tracers such as $^{12}$CO, $^{13}$CO and C$^{18}$O. The filament shows
a large $\sim 10$~km~s$^{-1}$ change in radial velocity over its
length, with the south end (the ``tail'') blue-shifted compared to the
north end. Also, based on 3D dust mapping, \citet{schlafly2015} found
evidence that the tail is several tens of pc more distant than the
northern end. Unless the Orion A cloud is stretching out along its
orientation in plane of the sky---which is unknown because of the lack
of accurate proper motion measurements---this would suggest that the
Orion A filament is compressing along its length. As for stellar
kinematics, \citet{furesz2008} and \citet{tobin2009} conducted an
optical survey measuring $v_r$ in the northernmost third ($\sim 2\deg$
long) of the region, centered on the ONC. In this region, they found
that young stars' velocities follow the gas ($^{13}$CO) velocity,
although their distribution is slightly asymmetric with a broad
blue-shifted tail not present in the gas component. They also noted
that the large scale $v_r$ gradient steepens north of the ONC, which
had been interpreted as evidence for large scale infall.

In our IN-SYNC survey, we obtained 4828 spectra of 2691 individual
sources throughout the $\sim 6\deg$ long region. The spatial
distribution of these sources is shown in Figure \ref{figure:radec},
left panel, together with a $^{13}$CO map from
\citet{nishimura2015}. In the previous paper of this series,
\citep[][,hereafter Paper IV]{dario2016}, we used these data to study
the properties of the stellar population. We focused on the fitted
stellar parameters ($T_{\rm eff}$, $A_V$, $\log g$, $v_r$), positioned
the sources in the Hertzsprung-Russel diagram (HRD), and assigned
stellar ages.

{ Our stellar sample was limited to $H<12.5$~mag, and assembled prioritizing known members from a multitude of literature estimates; additional sources with unknown membership were then added if in the luminosity range reached by our survey.
Our sample reaches masses as low as $M\sim0.15$~M$_{\odot}$ for small values of $A_V$, which in turn covers values up to $\sim20$~mag with a mean $A_V=2$~mag. The coverage of known members within the luminosity range of the survey is nearly complete throughout the region, with the exception of the central part of the ONC, where fiber collision constraints, combined with crowding, limited us to about 50\% completeness.
Stellar parameters are very accurate for cool stars ($T_{\rm eff}<5000$~K), with mean uncertainties in $T_{\rm eff}$ of the order of 50~K and mean and median errors in $v_r$ of $\sim1$ and $0.3$~km~s$^{-1}$. Our precision worsens at early spectral types; however, due to the IMF, only a minority of members are in this range.
}

We found a clear confirmation of a genuine age spread
throughout the region, from the correlation between HRD ages and
surface gravity-derived ages, as well as anticorrelation between ages
and IR excess, and ages and extinction. We also confirmed that the
non-embedded population of young stars around NGC1980 is older than
the rest of the system, but suggested that it is not part of a
well-separated cluster. This is because the kinematic properties of
these sources are indistinguishable from those of younger, more
embedded stars at the same position. Based on stellar parameters as
well as radial velocities, we found $\sim400$ new candidate
members. Most of these turn out to be young diskless sources, in
portions of the Orion A cloud not fully covered by previous studies to
assign memberships.

In this paper we continue the analysis of our IN-SYNC dataset, focusing on the kinematics and dynamics of the population.

\section{Kinematic comparison between stars and gas}
\label{section:gas-vs-stars}

\begin{figure*}
\plotone{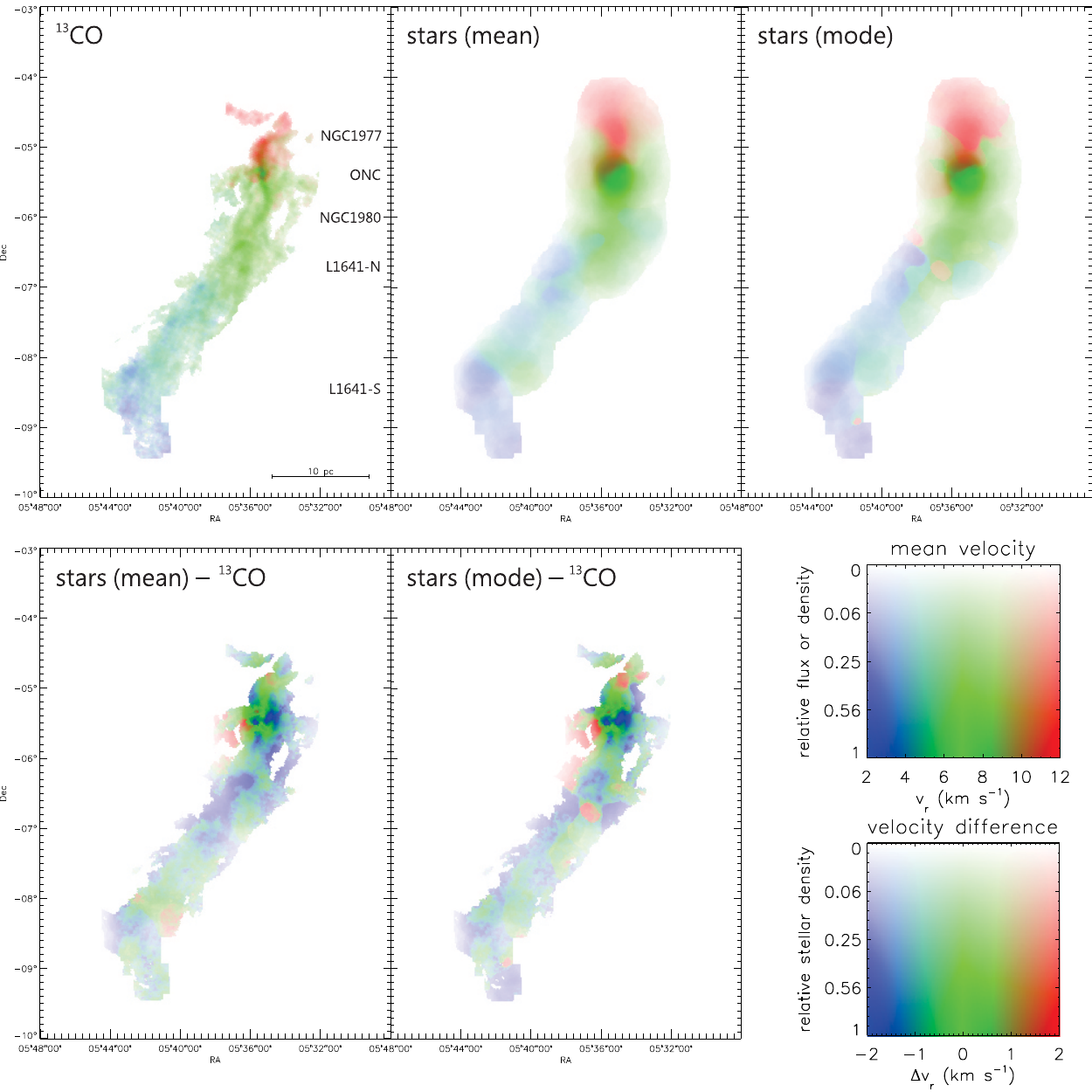}
\caption{Maps of mean or { mode} velocity for gas and stars, and difference between stars and gas, color-coded as indicated in the legend.  \label{figure:mean_vel_diff}}
\end{figure*}

Figure \ref{figure:radec} shows a map of the Orion region as well as a
position (Dec)--velocity diagram of our IN-SYNC targets in comparison
with that of the molecular gas tracers. Figure \ref{figure:3dvideo}
shows an animation of the full 3D (R.A., Dec and $v_r$) cube of the
data.  It is qualitatively evident that stars follow the global
motions of the molecular gas throughout the region, as seen before
limited to the north portion of the cloud ($-6\deg < \delta < -4\deg$,
\citealt{tobin2009}). Also, unsurprisingly, non-members tend to have
velocities that are different from the bulk of the young
population. Another feature is the fact that there appear to be more
members with $v_r$ lower than the peak value than stars with higher
velocities, in line with the results from \citet{tobin2009} that the
stellar velocity distribution is asymmetric with a broader
blue-shifted wing.

To investigate this possibility, we construct positional maps of mean
stellar velocities. First we exclude from our analysis all
non-members; we also exclude sources with velocities strongly
different from that of the bulk of the population ($v_r < -20$~km
s$^{-1}$ or $v_r>35$~km s$^{-1}$), to exclude outliers that are either
erroneous membership or velocity assignment, binary systems, or
ejected members. Then, for every angular position in $(\alpha ,
\delta)$ we consider the sources within a circular aperture with
radius of $20\arcmin$, provided that there are at least 30 sources in
the aperture, and we compute the weighted average of their $v_r$
values, as well as the median and mode value. Lastly, we smooth each
map with a kernel of $5\arcmin$ of radius.

\begin{figure*}
\epsscale{1.1}
\plotone{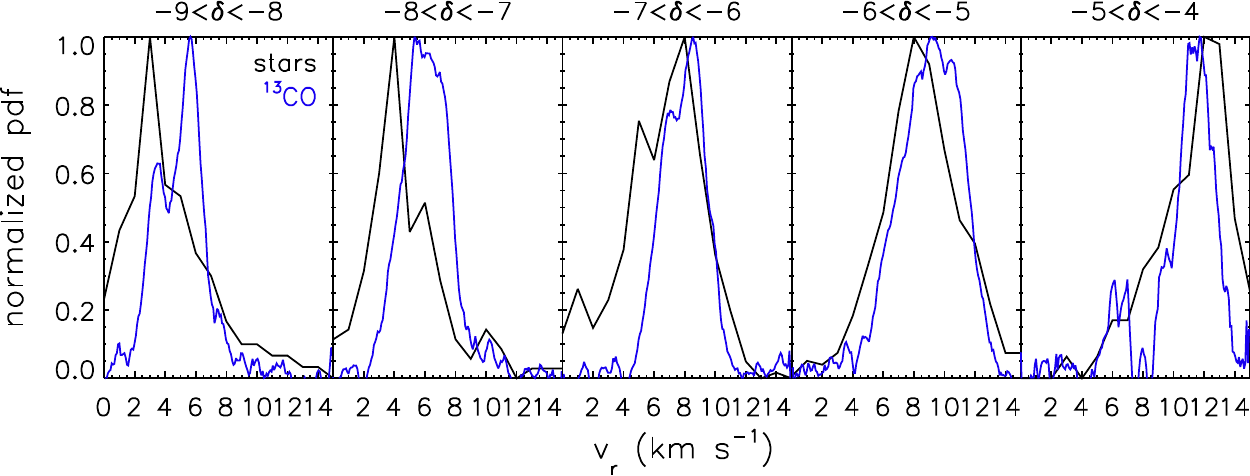}
\caption{The $v_r$ distribution for stars and gas ($^{13}$CO), divided into bins in declination. \label{figure:rv_distro_sliced}}
\end{figure*}

Figure \ref{figure:mean_vel_diff} shows the derived maps, as well as
the difference between stars and gas, color-coded according to $v_r$
and density or flux. Visually, it appears that the difference stars
$-$ gas, which will be referred to as $\Delta v_r$, is on average more
negative then positive, as the overall color leans toward the blue. In
fact, the area-weighted mean value of $\Delta v_r$ using the stellar
mean velocity is $-0.87$~km~s$^{-1}$, which decreases { in absolute value} to
$-0.44$~km~s$^{-1}$ weighting the result on the stellar density. These
values decrease, respectively, to $-0.66$~km~s$^{-1}$ and
$-0.25$~km~s$^{-1}$ when adopting the mode of the stellar
velocities. Such difference is also evident from Figure \ref{figure:rv_distro_sliced}, where we show the velocity distributions of stars and gas in different declination bins. Therefore, it appears that stars are systematically
slightly blue-shifted compared to the gas, and this is in part due to
an asymmetry in the $v_r$ distribution, as manifested by the
difference between mode and mean $v_r$.  \cite{}

\begin{figure}
\epsscale{1.1}
\plotone{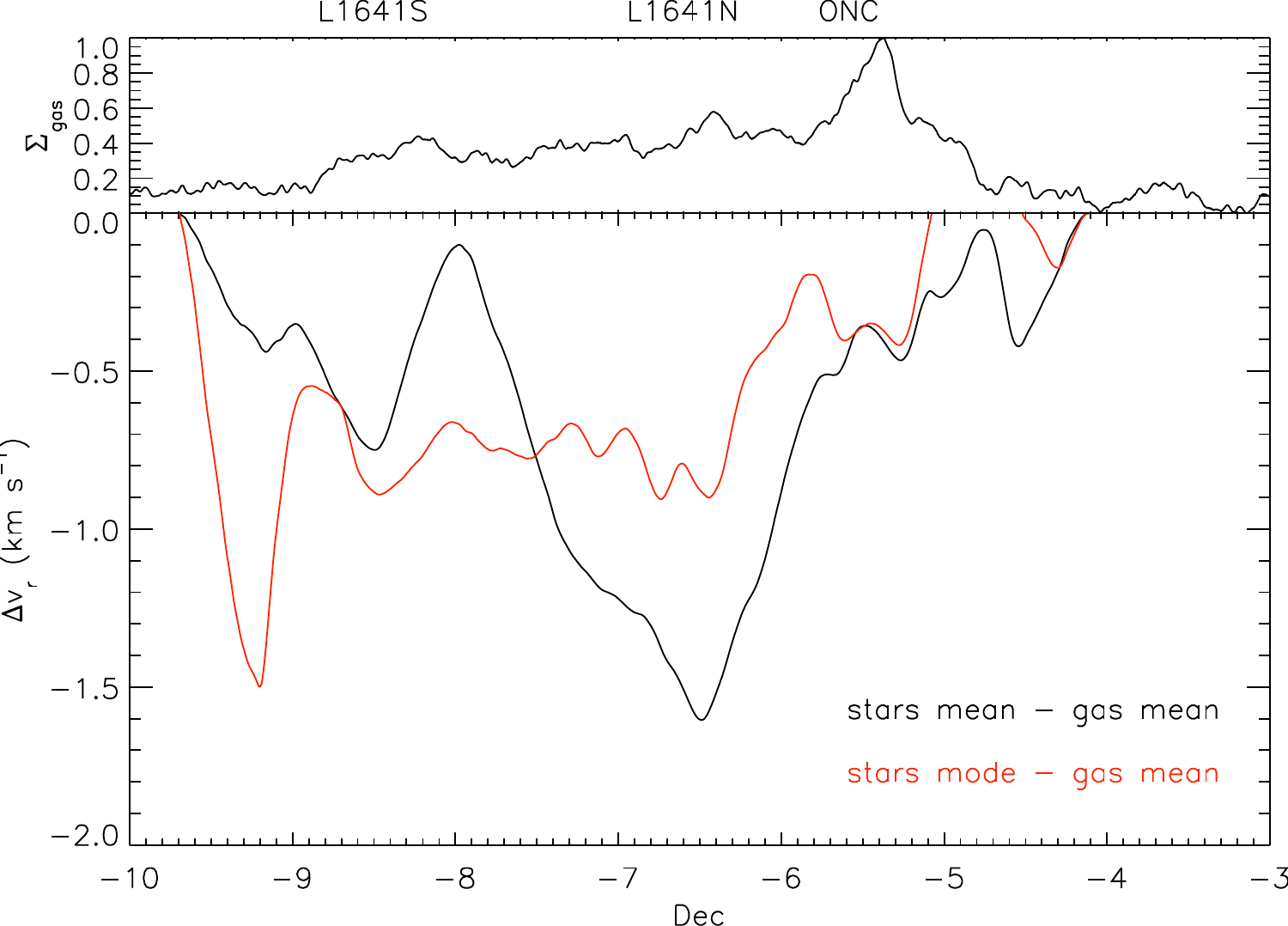}
\caption{Variation of $\Delta v_r$ -- stars-gas ($^{13}$CO) -- versus declination, using the mean or the mode of the stellar $v_r$ distribution as indicated in the legend. \label{figure:deltav_dec}}
\end{figure}

It is possible that this offset is not real, and simply due to
zero-point offsets in $v_r$. However, by comparing our radial
velocities with those from \citet{tobin2009}, for 550 matching
sources, we find a difference in the values of
$0.11\pm0.05$~km~s$^{-1}$, much smaller than the offsets between stars
and gas. Unless both, despite using different instrument and
techniques, have a similar systematic error in $v_r$, this suggests
that our IN-SYNC velocities are well calibrated.


Figure \ref{figure:deltav_dec} shows the dependence of the $\Delta
v_r$ on position in declination, computed from the maps of Figure
\ref{figure:mean_vel_diff} weighting, in Right Ascension, on stellar
density.  The largest offset is found in the central part of the
cloud, at $-7\deg\lesssim\delta\lesssim-6\deg$, around L1641-N and
NGC1980. Here the mean and mode of the velocity distribution have the
largest difference, suggesting a very asymmetric distribution. As
anticipated, and discussed in Paper IV, this part of the Orion A
stellar population is $\sim40\%$ older than the ONC and L1641-S and is
far less extincted suggesting that it is slightly in the foreground
and less associated with the molecular material. A bias due to
incompleteness at high $A_V$ is not the case, since the column density
of the molecular material in this region is relatively low, up to a
few magnitudes in $A_V$ \citep{lombardi2014}. Thus, star formation in
Orion A initiated earlier in the central part of the cloud, which is
closer along the line of sight that the remaining gas in the same
direction.

An offset in $v_r$ between stars and gas may have different
origins. One possibility is that the central part of the cloud
presented a primordial substructure along the line of sight, both
spatially and kinematically. Gas which formed the older population of
L1641-N and NGC1980 was closer and with small 1~km~s$^{-1}$
blue-shifted motion than the remaining gas today. Such offset is in
line with the typical turbulent motions within a molecular cloud
\citep[e.g.][]{hennebelle2012}. However, comparison of mean velocities
of $^{12}$CO(2-1) and $^{13}$CO(2-1) (expected to trace lower and
higher density regions, respectively), finds only modest differences
of $\sim0.1$~km~s$^{-1}$. However, in studies of Infrared Dark Cloud
(IRDC) filaments, \citet{henshaw2013} have found offsets of similar
magnitude between denser gas traced by N$_2$H$^+$(1-0) and lower
density gas traced by C$^{18}$O(1-0).

Another possibility is that prolonged stellar feedback of this
foreground population pushed back on the remaining background gas,
which is now receding 1~km~s$^{-1}$ with respect to the stellar
population.

If primordial substructure along the line of sight is present, the
asymmetry of the velocity distribution could also originate from the
superposition of two kinematically separated substructures of the
cloud.  Since, as shown in Paper IV, we found that the $v_r$
distribution around L1641-N does not vary with stellar age, these structures would have needed to
sustain star formation at the same time, and with similar duration,
which seems somewhat unlikely.

\section{Kinematic subclustering}
\label{section:subclustering}

The presence of kinematic substructures in the stellar population of
Orion A, and their comparison with those of the gas, can provide
important clues on the primordial substructure of the parental
star-forming gas. On the other hand, smooth distributions in space and
velocity could indicate that dynamical evolution has occurred, since
spatial and kinematic substructure is expected to be erased over time
in a gravitationally bound system. Evidence of this has been found in
the ONC \citep{dario2014b}, where the population in the denser, more
dynamically evolved core has less spatial substructure in comparison
to the cluster outskirts.

An analysis of the kinematic substructure in
position-position-velocity ($ppv$) in Orion A was presented by
\citet{hacar2016}, using a preliminary version of our IN-SYNC
data. These authors adopted a \emph{friends-of-friends} algorithm that
isolates groups or chains of sources in which the separation of each
member of the group, in the $ppv$ space, from another member of the
group is smaller than a given threshold. Imposing groups with at least
4 sources, they find 37 of such groups, most of them (30) in the
northern region of the cloud (L1461N and above), significantly more
than expected from random positions in $ppv$ space.

{ There is no optimal method to identify structures of discrete points in a 3-dimensional space, rather, different techniques can be developed depending on what one defines as a ``structure''.
In Sections \ref{section:subclustering:peaks} and \ref{section:subclustering-discrete} we describe two separate methods and discuss the results.
\begin{itemize}
  \item \emph{ppv Peaks}: this technique is based on building a 3-dimensional density map in $ppv$ space from our discrete stellar sample. On this map, then, peaks are identified as overdensities in $ppv$ space. Individual stars can be associated to one or another peak. This method is somewhat analogous to the algorithm \texttt{Clumpfind} \citep{williams1994}, widely used to identify $ppv$ structures in continuous molecular line spectral cube data. This method is sensitive to centrally concentrated overdensities in $ppv$, but does not necessarily identify filamentary or very elongated structures.
  \item \emph{Connected Structures}: this technique does not adopt a density map in $ppv$, but rather builds groups of stars in $ppv$ starting from the highest local density of stars in $ppv$ and adding individual stars as nearest neighbors until a given density threshold is encountered, before creating a new separate group. While this technique does not require kinematic structures to be centrally concentrated, the number and size of structures can be very sensitive on small perturbations of the position of individual sources in $ppv$ space.
\end{itemize}
}

\subsection{$ppv$ Peaks}
\label{section:subclustering:peaks}

To construct stellar density maps in $ppv$ space, we
consider each point of the $ppv$ space and find the closest $n$ stars,
in $ppv$ to that point. We adopt a conversion metric such that $1\deg$
in spatial distance in RA and Dec corresponds to 4~km~s$^{-1}$ in
$v_r$. This has been chosen so that the overall spatial extent of the
region (i.e., in declination) is comparable with the broadness of the
$v_r$ distribution. The stellar density associated to the point is
then $n/V$, where $V$ is the volume of the sphere of radius equal to
the distance to the $n$-th star. Therefore, our density maps naturally
trace the density of groupings of stars at a given number scale
defined by $n$.  Lastly, coherent structures in $ppv$ space are
identified as local density maxima on these density maps within a
kernel size of 10\arcmin or 0.66~km~s$^{-1}$. For each structure, we
also measure the density $\rho(\alpha,\delta,v_r)$, as well as the
density ``contrast'' $\delta\rho$ against the local background in
$ppv$ space, for which we adopted the density maps obtained for
$n_{\rm bg}=30$.

\begin{figure*}
\epsscale{0.85}
\plotone{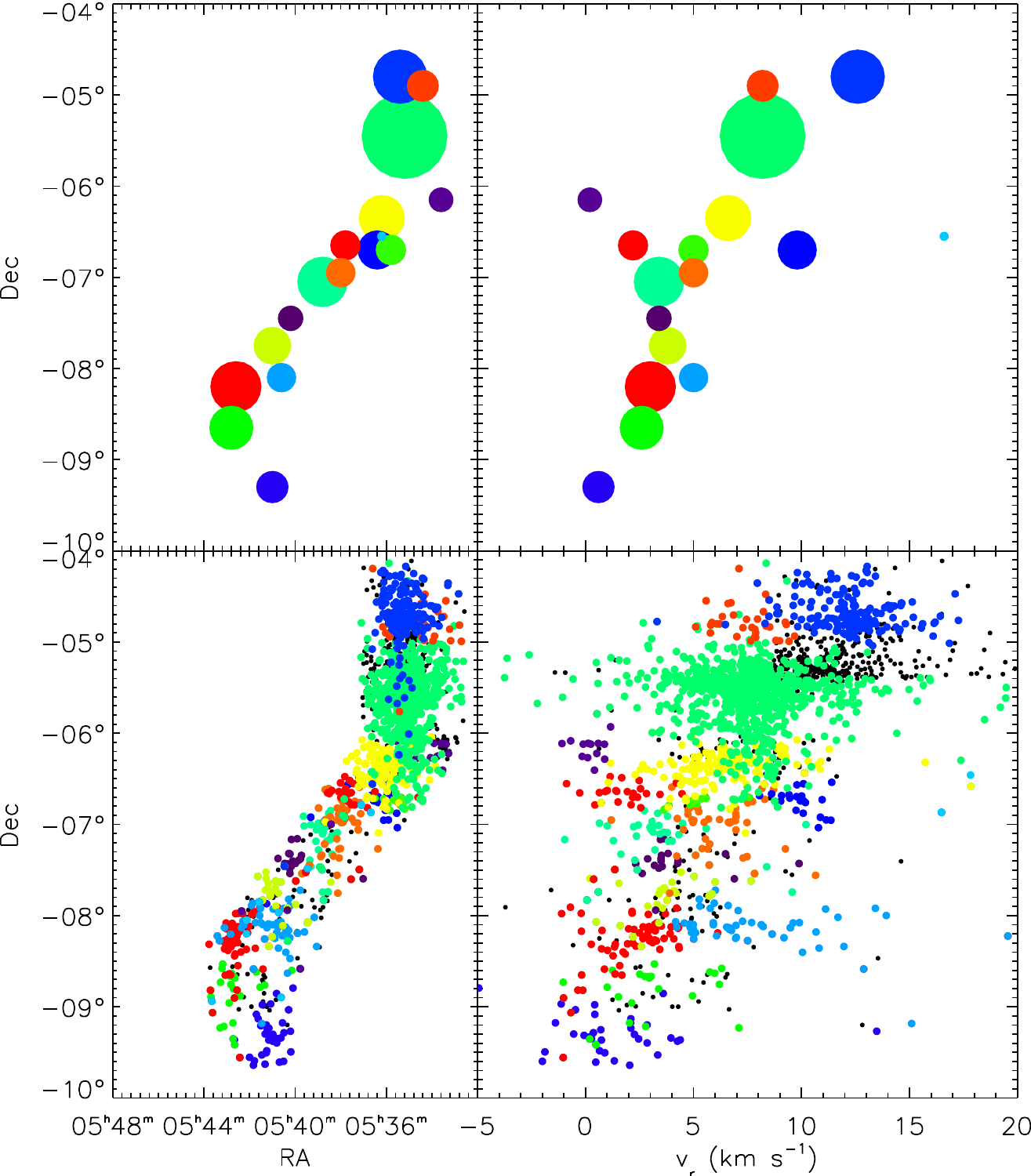}
\caption{
{\em Top panels:} position-position and position-velocity diagrams of
the identified overdensities in $ppv$ space. The size of the circles indicates the
the local density of the structure in $ppv$, while the colors are arbitrarily chosen to easily distinguish between different groups. {\em Bottom panels:} position of the
members in the same projection of the $ppv$ space. Colored dots
indicate stars associated in $ppv$ to each of the identified
substructures (see text), with colors corresponding to each structure in the upper panels.  \label{figure:ppv_subclustering5}}
\end{figure*}

\begin{table}[]
\caption{Substructures in $ppv$ space}
\begin{center}
\begin{tabular}{r|rr|rr|}
    \hline
    \multicolumn{1}{r|}{ } & \multicolumn{2}{r|}{Measured} &  \multicolumn{2}{r|}{Random simulation}\\
    \multicolumn{1}{r|}{$n$} & \multicolumn{1}{r}{$N_{\rm structures}$} &  \multicolumn{1}{r|}{$\delta\rho$} & \multicolumn{1}{r}{$N_{\rm structures}$} &  \multicolumn{1}{r|}{$\delta\rho$}\\[1ex]
    \hline
    & & & \\
      4 &  20 & 16.347  & 14.760$\pm$2.619   & 1.405$\pm$1.336 \\
      5 &  17 & 10.740  & 13.920$\pm$3.081   & 1.184$\pm$1.243 \\
      6 &  14 &  9.215  & 12.360$\pm$2.797   & 0.937$\pm$0.729 \\
      7 &  14 &  6.486  & 11.880$\pm$2.862   & 0.899$\pm$0.75  \\
      8 &  13 &  5.427  & 10.520$\pm$1.475   & 0.851$\pm$0.537 \\
     10 &  12 &  3.431  &  9.920$\pm$1.605   & 0.702$\pm$0.406 \\
     20 &   8 &  2.050  &  7.440$\pm$1.635   & 0.500$\pm$0.163 \\[1ex]
     \hline
\end{tabular}
\end{center}
\label{table:ppv-substructures}
\end{table}

\begin{figure*}
\epsscale{0.85}
\plotone{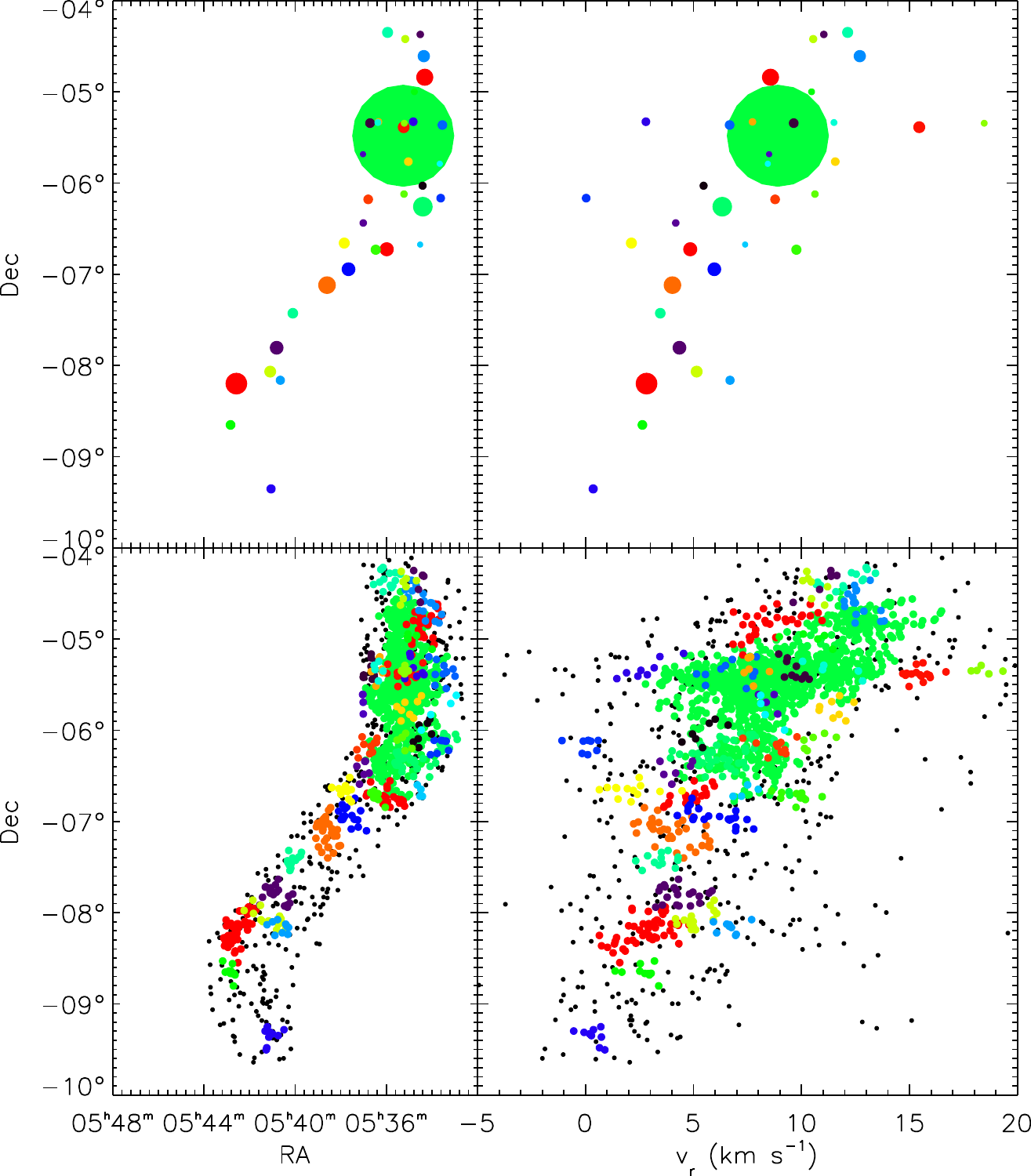}
\caption{
Similar to Figure \ref{figure:ppv_subclustering5}, but now showing an
example of the $ppv$ Connected Structures method, described in
\S\ref{section:subclustering-discrete}, in this case for
$\rho_0=0.3$~pc$^{-2}$(km~s$^{-1}$)$^{-1}$, $n=5$ and
$\delta_v=8$km~s$^{-1}$ ({\em see text}).
The top panels show the center of each structure, as the mean of positions and $v_r$ of the stars associated to it. The area of the
circles in the top panel is proportional to the number of stars in
each group, which in turn are shown in the bottom panels. The colors, corresponding between the different panels for each structure, are arbitrary to
facilitate the visual distinction between structures.  \label{figure:ppv_subclustering5_discrete}}
\end{figure*}

Figure \ref{figure:ppv_subclustering5} (upper panels) shows the result
for $n=5$. { It is
noteworthy that the majority of these groupings are located along the
filament south of the ONC, where
about half of our IN-SYNC members are concentrated. On the other hand ONC as a whole, accounting for almost half of the stellar population in the cloud,
is found to be a single kinematic substructure in $ppv$}. \citet{dario2014b} also showed the projected spatial
positions of young members in the ONC, especially the central regions,
have reached a relatively smooth, quasi-isotropic distribution, and
that the population within the half mass radius $r_h\lesssim1$~pc is
at least 7 free-fall times old. This evidence, together with the lack
of kinematic substructure we have now identified, points to a
relatively old dynamical age for the ONC \citep{tan2006}, suggesting
that primordial structures in space and velocity may have been erased
by dynamical interactions. This is in agreement with estimates that
the ONC is at least 4 free-fall times older within the half-mass
radius $r_h\simeq1$~pc \citep{dario2014b}.

{ Figure \ref{figure:ppv_subclustering5} also shows that at the location
of L1641N and NGC1980 ($-7.5\deg\lesssim\delta\lesssim-6\deg$)
substructure is significant: the different substructures show a large difference in velocity, more than elsewhere. In this region, as shown in Section \ref{section:gas-vs-stars}, the overall stellar velocity distribution is asymmetric, thus this asymmetry is likely caused by the superposition of kinematically distinct groups of stars in the same region of the cloud.
}

Having identified our kinematic substructures as density peaks in $ppv$, we can attempt to associate individual stars to each substructure. The criterion we chose is the following: a star belongs to a kinematic substructure if, along a segment in $ppv$ space spanning from the star to the density peak, the $ppv$ density increases monotonically. An illustration of the result is shown in Figure \ref{figure:ppv_subclustering5}, bottom panels.

Increasing $n$ decreases the overall number of identified
substructures in $ppv$ space, as individual small groupings of stars
merge into larger single structures (see Table
\ref{table:ppv-substructures}). For $n=20$, the only structures
detected are 3 groups for L1641S, 2 for L1641N (all roughly coinciding
to the densest groupings of stars in R.A.--Dec space, plus two
structures in the ONC region, and one north of it.

We performed tests to evaluate the statistical significance of our
detected kinematic substructures, in particular for what concerns the
velocity axis of $ppv$ space. To this end, we created artificial
stellar distributions in $ppv$ space, where the positions are the same
as our IN-SYNC members, and the velocities are randomly drawn from the
local $v_r$ distribution. This latter was computed, for each position
in $(\alpha,\delta)$ measuring the weighted $v_r$ mean of the nearest
30 members, as well as the measured dispersion, after one iteration of
sigma clipping, with a $3\sigma$ threshold, to remove outliers. From
each realization of artificial distributions in $ppv$, density maps
are created and kinematic substructures isolated, with the same method
as that applied on the actual data. For each value of $n$, 50 of these
artificial tests have been performed, and the results are summarized
in Table \ref{table:ppv-substructures}. We find that, regardless of
$n$, the number of substructures we detect is compatible with that
from random artificial experiments. This, however, is in part due to
the fact that our artificial tests kept the stellar positions, and
randomized only the velocities. Therefore spatial groupings of sources
in the Orion A may still lead to the detection of structures in $ppv$
in spite of a large degree of randomness in $v_r$.  On the other hand,
the density contrasts $\delta\rho$ resulting from our data are
significantly higher than those from artificial $ppv$ distributions;
this confirms that the kinematic substructures we identify in $ppv$,
have a statistically significant higher coherence in velocity, when
compared to the ``local'' velocity distribution of the nearest 30
stars for each position in the Orion A cloud.

\subsection{$ppv$ Connected Structures}
\label{section:subclustering-discrete}

The method described in \S\ref{section:subclustering:peaks} is
particularly sensitive to structures that appear as centrally
concentrated ``clumps'' in $ppv$ space; coherent structures with more
irregular shapes may not be detected. Thus, we introduce a
conceptually different approach and analyse the different findings. In
this case, we do not build a 3-dimensional map of stellar density in
$ppv$, but we consider the stellar positions in such space as discrete
elements. We associate to each star a local density in $ppv$, as above
by measuring the volume containing the closest $n$ stars. We then
begin by considering the star with the highest density peak, and use
this as the starting point for a group of stars in $ppv$. We expand
this group by adding iteratively one star at a time in order of
closest distance in $ppv$ space from any of the group members.
The growth of the group halts when the next star about to be added
falls at a density below a fixed threshold, $\rho_0$. Next, a new
group is started from the highest density star among those not yet
included in a previous group, assuming it is above this threshold.

We vary three parameters: $n=5$ or 10, $\rho_0=1$ or
$0.3$pc$^{-2}$km$^{-1}$s$^2$, and also the metric to measure the
distance in $ppv$ as the interval $\delta_v$ on the $v_r$ axis, in
km~s$^{-1}$, corresponding a spatial interval of 1\deg; in this case
we assume $\delta_v=4$, as in \S\ref{section:subclustering:peaks}, but
also test values of 2 and 8. An example of the results is shown in
Figure \ref{figure:ppv_subclustering5_discrete}. Overall, we find a
higher number of structures compared to the method of
\S\ref{section:subclustering:peaks}, as we are able now to identify
smaller structures that would not have been resolved given the kernel
size in $ppv$ we previously adopted. As a result of this, while the
ONC remains the largest identified structure, { some additional smaller substructures are now detected in its vicinity}. As with the identifications of peaks in the $ppv$ density
maps, even in this case we find the L1641 region to be highly
substructured. Decreasing $\rho_0$ increases the fraction, $f$, of the
total stellar sample included in some identified $ppv$ groups;
however, we do not detect a clear difference in the total number of
groups. No significant difference in the results is found changing $n$
or $\delta_v$.

\begin{table}[]
\centering
\caption{Groups of stars identified in $ppv$ space and comparison with random tests. See text.}
\label{table:ppv-substructures-discrete}
\begin{tabular}{lll|ll|ll|ll}

$\rho_0$             &           $n$       & $\delta_v$ & $N_{\rm groups}$ & $\sigma_{N_{\rm groups}}$  & $f$ & $\sigma_f$ & $p1$ & $p2$  \\ \hline
\multirow{6}{*}{1}   & \multirow{3}{*}{5}  & 2 &   36  &   0.4 &  30.22  &  0.87  & 0.285  & 0.372 \\ \cline{3-9}
                     &                     & 4 &   35  &   0.3 &   31.2  &  0.79  &  0.84  & 0.105 \\ \cline{3-9}
                     &                     & 8 &   36  &   0.6 &   30.5  &  0.76  & 0.291  & 0.204 \\ \cline{2-9}
                     & \multirow{3}{*}{10} & 2 &   27  &  -0.5 &  38.19  & -0.76  & 0.329  & 0.626 \\ \cline{3-9}
                     &                     & 4 &   25  &  -0.7 &  41.52  & -0.53  & 0.183  & 0.003 \\ \cline{3-9}
                     &                     & 8 &   23  &  -1.2 &  45.26  & -1.11  & 0.019  & 0.533 \\ \hline
\multirow{6}{*}{0.3} & \multirow{3}{*}{5}  & 2 &   33  &   2.6 &   47.0  &  2.71  & 0.003  & 0.071 \\ \cline{3-9}
                     &                     & 4 &   30  &   2.2 &  51.73  &  1.93  & 0.173  & 0.245 \\ \cline{3-9}
                     &                     & 8 &   36  &   3.3 &  43.42  &  1.94  & 0.007  & 0.018 \\ \cline{2-9}
                     & \multirow{3}{*}{10} & 2 &   29  &   2.8 &  52.62  &  1.88  &   0.0  & 0.008 \\ \cline{3-9}
                     &                     & 4 &   34  &   4.7 &  45.12  &  1.98  & 0.002  & 0.001 \\ \cline{3-9}
                     &                     & 8 &   31  &   3.2 &  49.87  &  1.68  &   0.0  & 0.009 \\ \hline
\end{tabular}
\end{table}

As before, we run the algorithm to artificially generate distributions
in $ppv$ space (see \S\ref{section:subclustering:peaks}), and
investigate different diagnostics to test the significance of the
results. This is summarized in Table
\ref{table:ppv-substructures-discrete}. For each parameter choice we
compare the number of groups $N_{\rm groups}$ isolated from our data
with the mean and standard deviation of this quantity from the random
experiments; $\sigma_{N_{\rm groups}}$ in this context is the measured
excess, in standard deviations, from the result from random
experiments. Similarly, $\sigma_f$ is the departure in
standard deviations for the fraction of stars belonging to some
structure in $ppv$. Lastly, we use Kolmogorov-Smirnov test to verify
if two cumulative distribution functions are significantly different
between our measured values and those from randomly generated
distributions. Specifically we list $p_1$ as the KS probability (where
a low number indicates significant difference) of the distribution of
peak local densities in $ppv$ of the identified groups, and $p_2$ that
of the distribution of number of stars in the detected $N_{\rm
  groups}$. Results show that for high values of $\rho_0$, the
statistical significance is relatively poor or absent. Decreasing
$\rho_0$, thus enabling a larger number of sources to be labeled as
clustered, especially in regions of low density such as L1641, leads
to higher significance. Specifically, we identify a number of $ppv$
structures 2--4$\sigma$ larger than in random experiments, a fraction
$f\sim2\sigma$ higher, and probabilities $p_1$ and $p_2$ less than 1\%
in most cases.

\begin{figure}
\epsscale{1.1}
\plotone{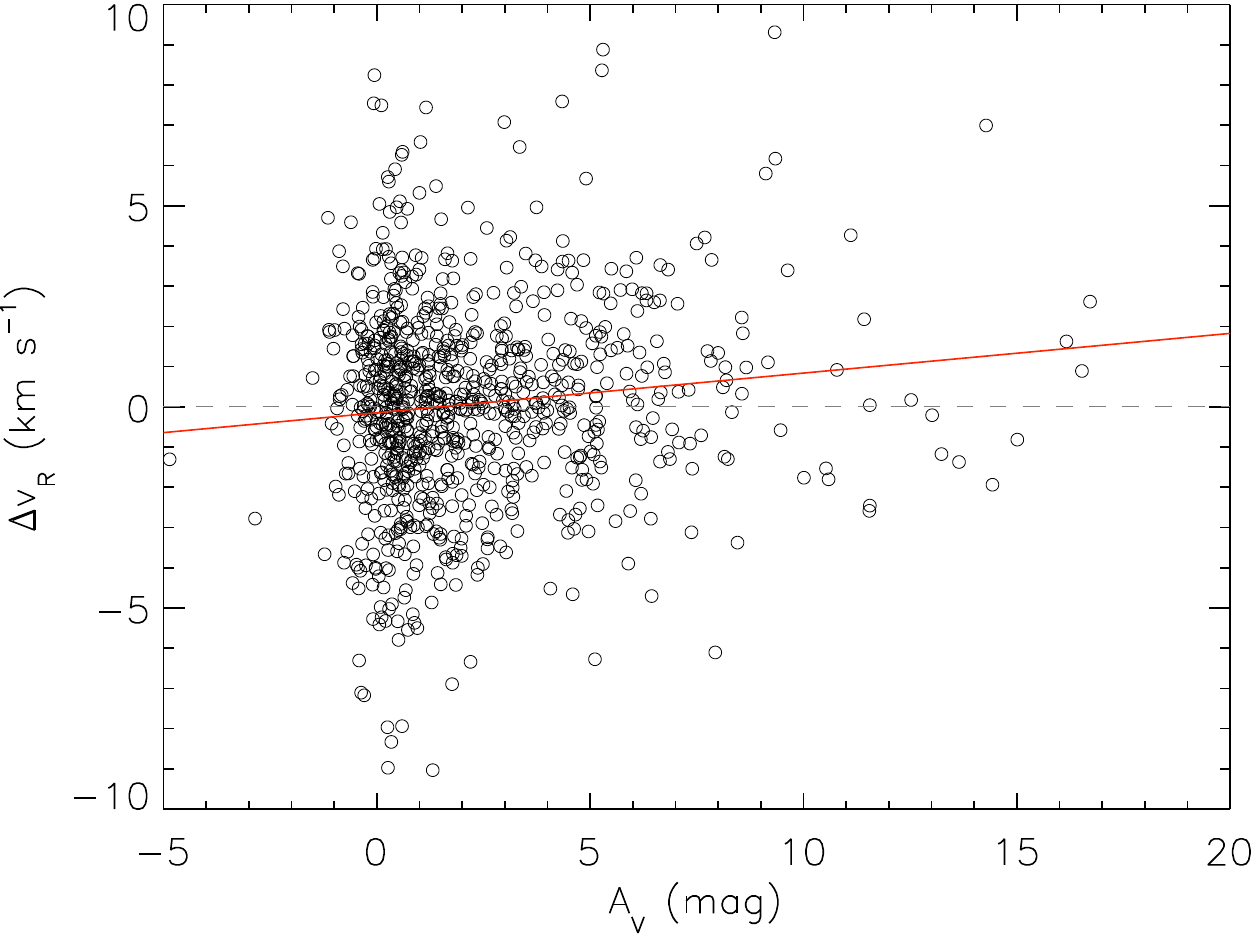}
\caption{
Dependence of the $v_r$ offset from the local mean $v_r$ with
extinction $A_V$, for a circular aperture of $1\deg$ in diameter
centered on the ONC, limited to members with $T_{\rm eff}<5000$~K. The
red line is the linear fit through the
data. \label{figure:av_deltarv}}
\end{figure}

\begin{figure*}
\plottwo{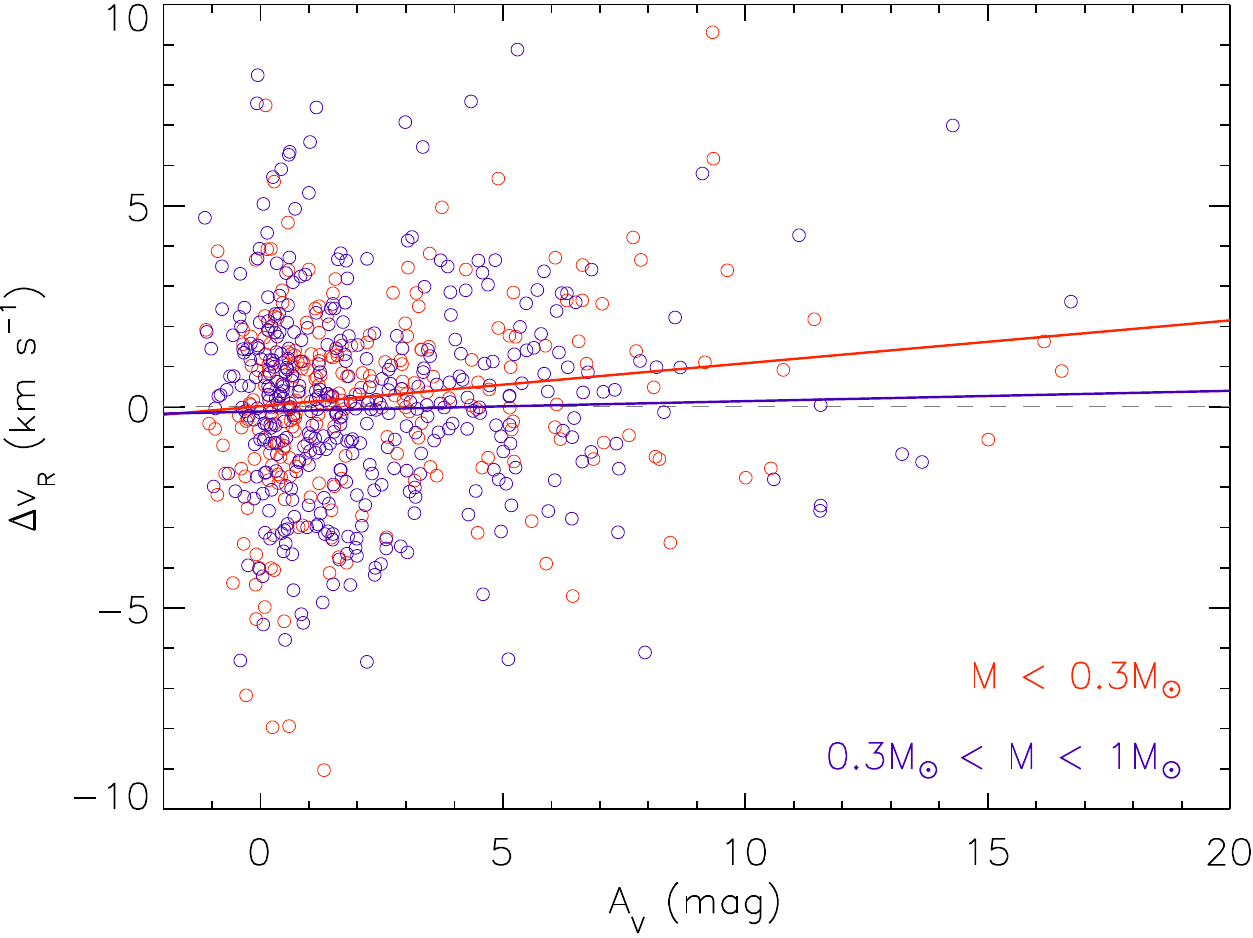}{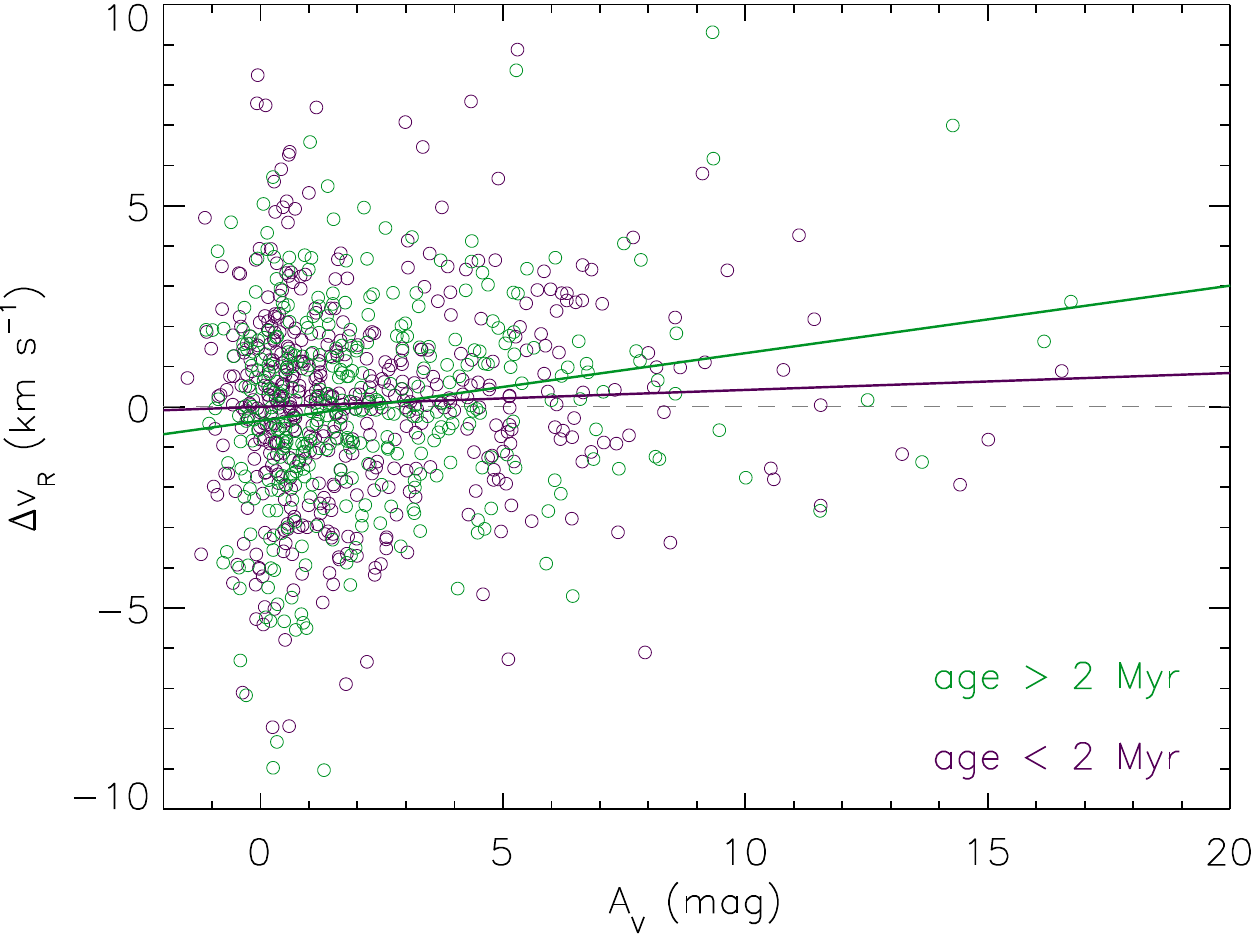}
\caption{
Same as Figure \ref{figure:av_deltarv}, but dividing the population
into ranges in masses (left panel) and ages (right
panel). \label{figure_av_deltarv_mass_age}}
\end{figure*}

The fact that we detect statistically significant overdensities in
$ppv$ space adopting two conceptually different methods
(\S\ref{section:subclustering:peaks} and
\S\ref{section:subclustering-discrete}) is highly indicative that
kinematic substructure in Orion remains present, and thus may be a
tracer of the original turbulent motions of the molecular material in
the parental cloud.  Future observations (i.e., multi-epoch) will help
improve the $v_r$ measurements, especially enabling the identification
of binaries. Proper motion measurements from the final {\it GAIA}
survey will be also useful to study kinematic substructures in
Orion. The degree of kinematic substructure that we measure will also
be compared to the results from theoretical/numerical models of star
cluster formation under different scenarios \citep{wu2017a,wu2017b}.

\section{Expansion of the ONC}
\label{section:expansion}

Limited to the ONC region of the system, we find evidence for a
correlation between radial velocities and extinction, in that sources
with higher $A_V$ are statistically more red-shifted compared to less
embedded members. This is shown in Figure \ref{figure:av_deltarv},
where the offset $\Delta v_r$ from the local mean radial velocity { (computed as in Section \ref{section:gas-vs-stars} and shown in Figure \ref{figure:mean_vel_diff}) }is
plotted against $A_V$, for all members within a circular aperture of
$1\deg$ centered on the Trapezium region that defines the center of
the ONC \citep[e.g.,][]{dario2014b}. We also restricted our sample to sources
with $T_{\rm eff}<5000$~K, as for higher temperatures the stellar
parameters---hence $A_V$, are more uncertain. A linear fit through the
data shows an increase in $v_r$ of about $\sim$0.1~km~s$^{-1}/A_V$. A
randomization test through the permutation of the values along the
$y-$axis of this diagram showed that this correlation is significant
at a $\sigma=3.1$ level. Both the correlation and its significance
level remain unchanged varying the aperture diameter between 0.5 and
2$\deg$; the significance reduces at smaller apertures because of low
number statistics, and at larger distances outside the ONC.

Since $A_V$ is a tracer of the depth of a star along the line of
sight---even though the distribution and volume density of the dust is
spatially inhomogeneous---this result suggests a global expansion
along the line of sight. We exclude that this correlation may be a
spurious outcome of correlated systematic effects in the extraction of
$v_r$ and $A_V$, since in \citet{cottaar2015} an opposite trend is
observed in IC~348 using APOGEE data and identical data reduction and
extraction of parameters as in this work. From the relation $\Delta
v_r/\Delta(A_V)\sim$0.1~km~s$^{-1}/A_V$ it follows that the central
90\% of the population along the line of sight, with $0\lesssim
A_V\lesssim 7.4$ shows an overall $v_r$ spread of
$\sim0.7$~km s$^{-1}$, or a net maximum expansion from the cluster's
midplane of $\sim0.35$~km s$^{-1}$. The reddening distribution is
highly asymmetric, with a median $A_V$---which could be associated
with the cluster's midplane---of 1.3~mag. Note that this fact is not
only seen in our IN-SYNC survey, but all previous studies reaching
higher completeness and $A_V$ \citep[e.g.]{dario2014b,robberto2010};
thus it is not the result of completeness limits of this work.  This
implies that the volume density of the molecular material increases
along the line of sight. We are unable to explicitly measure the net
velocity offset of two halves of the population from this midplane, as
in the low-extinction limit, we do not measure any correlation of
$v_r$ with $A_V$. This, however, is likely due to the fact that the
typical uncertainty in $A_V$ is $\sim0.4$ at all extinctions. Thus,
for low $A_V$ the relative uncertainty is particularly high,
flattening any $v_r$--$A_V$ correlation due to regression dilution.

We look for variations in the measured correlation indicating
expansion with mass and age, and divide the stellar samples in two
mass ranges ($M<0.3M_\odot$ and $0.3M_\odot<M<1M_\odot$) and two age
ranges on both sides of the median cluster ages. These parameters have
been obtained from \citet{siess2000} evolutionary models (Paper
IV). Although there is a hint that lower-mass stars show a large velocity
gradient with $A_V$, { and so do older stars compared to younger
ones} (see Figure \ref{figure_av_deltarv_mass_age}), these differences
are not statistically significant ($\sim 1\sigma$). Future data are
needed to confirm or not a mass and age dependence of the expansion of
the ONC.

Finally, no correlation between radial velocity and extinction is
found along the filament south of the ONC (L1641-N and L1641-S).

\section{Velocity dispersion and virial state}
\label{section:vel_dispersion}

We obtained the stellar velocity dispersion $\sigma_v$, for different
samples of the Orion A population, in a similar fashion as in Paper II
and III for NGC~1333 and IC~348, respectively, in
Perseus. { Specifically, we consider the observed distribution of $v_r$, as well as the measurement uncertainties in radial velocity, and the contribution from unresolved binarity. }
 Unlike the IN-SYNC survey in Perseus,
our $v_r$ values have only one epoch for the majority of our stellar
sample, therefore our binarity correction has been performed
statistically. We adopt a flat mass ratio distribution
\citep{reggiani2013}, a flat eccentricity distribution
\citep{duchene2013} over the whole mass range, and a log-normal period
distribution from \citet{raghavan2010}. Using the tool \textsc{VELBIN}
(Paper II), then, we generated a large number of binary orbits
randomly drawn from these parameter distributions, and assuming random
orientations we obtained the 1-dimensional distribution of radial
velocities of the primary. We assumed a binary fraction $f_{\rm bin}=0.44$,
as determined by \citet{raghavan2010} for solar-type stars.

{ Considering a sample of $v_r$ and associated errors $\delta v_r$, we generate a large number of intrinsic velocity dispersions $\sigma_v$ as normal distribution of varying standard deviation. We then convolve these with the contribution due to measurement errors; to this end we add together the individual distributions of errors, all assumed normal, for each star. Since stars have different errors, the resulting distribution is not a gaussian. Last, we consider the 1-dimensional distribution of $v_r$ offsets from binaries for $f_{\rm bin}=1$, scale it down to our value of $f_{\rm bin}$, complement it with $(1-f_{\rm bin})$ zeros representing single stars, and convolve it to the intrinsic $\sigma_v$ plus measurement error distributions. The resulting simulated distributions, as well as the measured $v_r$ distribution are fitted with gaussian distributions after removing outliers at $3\sigma$, and the best match is found. This naturally represents the best velocity dispersion $\sigma_v$ which, convolved with contributions from errors and binaries, reproduced the observed $v_r$ distribution. An example of this procedure is shown in Figure \ref{figure:sigmafit_example}, for a sample of members centered on the ONC within a radius of $0.3\deg$ or $\sim2.2$~pc.

{Generally, the contribution of the measurement errors is far smaller than both the observed $v_r$ distribution, and the intrinsic $\sigma_v$. For example the distribution of $v_r$ offsets due to measurement errors, shown in Figure \ref{figure:sigmafit_example} is such that 68\% of the data (i.e., equivalent to a 1$\sigma$ interval) is in the range $\pm0.38$~km~s$^{-1}$.
 Furthermore, as pointed out in Paper~III, the contribution from binary is even smaller (68\% of $v_r$ offsets in the range $\pm0.10$~km~s$^{-1}$ for $f_{\rm bin}=0.44$). Therefore our results are not very sensitive to the assumptions made for the binary population, i.e., the distribution of binary parameters and $f_{\rm bin}$.
This fact would be less the case for high-mass stars, both due to a higher binary fraction, and possibly a lower intrinsic $\sigma_v$ from dynamical energy equipartition. However, our sample is dominated by low-mass stars, with 95\% of members with a mass lower than 2~M$_\odot$.
 }

\begin{figure}
\epsscale{1.1}
\plotone{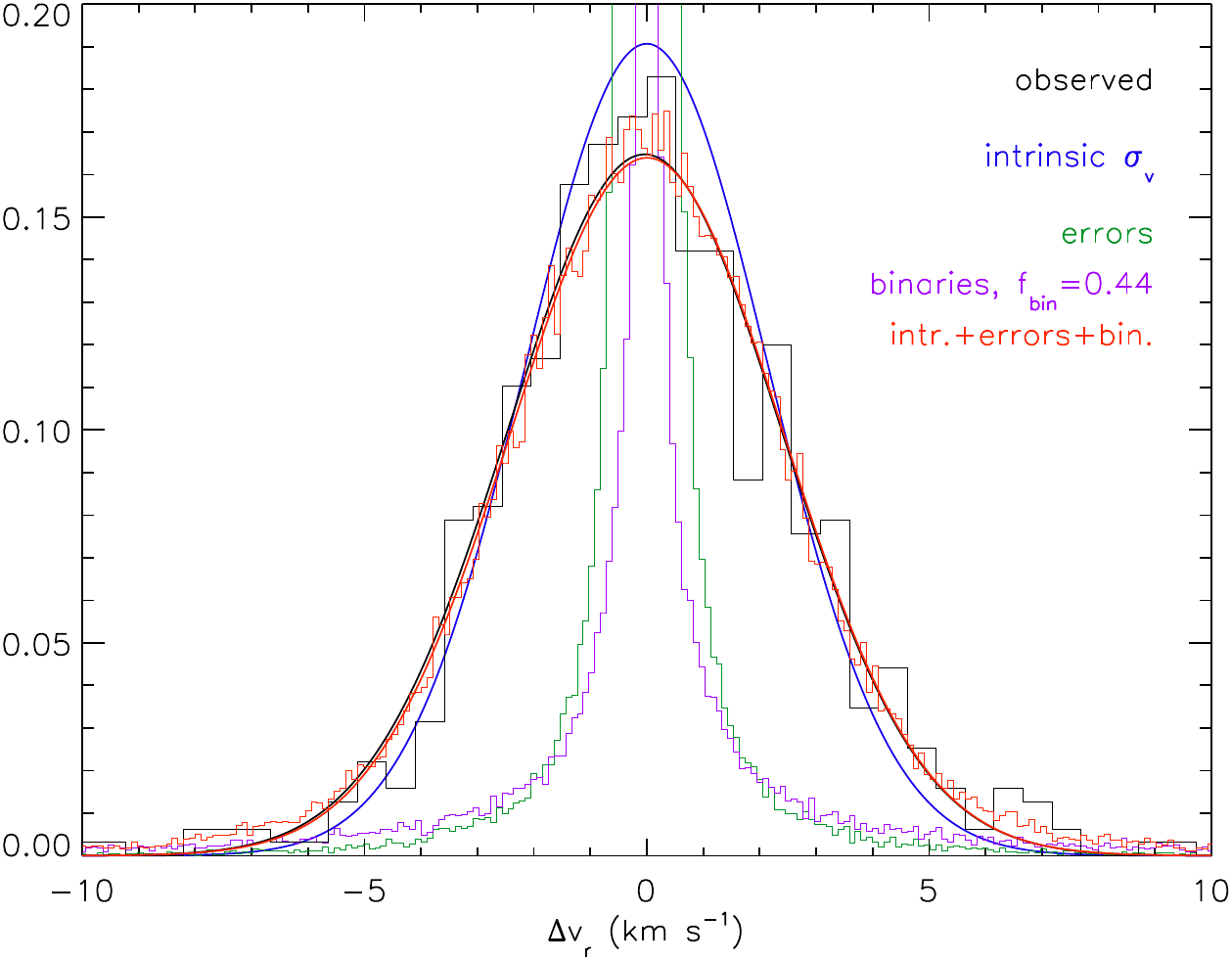}
\caption{
Example of the extraction of the velocity dispersion $\sigma_v$ from the observed $v_r$ distribution, correcting for $v_r$ uncertainties and unresolved binaries (see text). The distribution shown is for the ONC population, within a circle of $0.3\deg$ ($\sim2.2$pc) in radius. \label{figure:sigmafit_example}}
\end{figure}

Figure \ref{figure:dec_dispersion} shows the derived, corrected, $\sigma_v$ in
different bins in declination along the Orion A cloud. The solid line represents the
dispersion as is, adopting the measured $v_r$ for all stars in each
bin. We have also corrected for the spatial gradients in $v_r$ along
the region (see \S\ref{section:gas-vs-stars}), which if not accounted
for, may lead to an overestimate of the overall $\sigma_v$ in a
region. To this end we adopted the radial velocities as the
difference, star by star, between the measured $v_r$ and the local
mean radial velocity in that position, as computed in
\S\ref{section:gas-vs-stars} and shown in Figure
\ref{figure:mean_vel_diff}, second panel. We refer to these two
dispersions as ``standard'' and ``gradient-corrected''. The result is
shown in Figure \ref{figure:dec_dispersion} as a dashed line. { Overall,
$\sigma_v$ { varies in the range 1--2.5~km~s$^{-1}$.} The most striking feature is a broad peak
in $\sigma_v$ in L1641N; this is consistent with the detected
asymmetry of the velocity distribution and the presence of kinematic
substructure in this region, possibly indicating spatial structure
along the line of sight (see Section \ref{section:subclustering}). Thus a
larger $\sigma_v$ could be the result of superposition of stellar
systems with relative bulk mean motions. Just south of the ONC we detect the most prominent drop in velocity dispersion.  The ONC also reveals itself as a local peak in $\sigma_v$ at values 2--2.5~km~s$^{-1}$.
Moving north of the ONC, in the region of  NGC~1980 and NGC~1977/OMC 2-3, $\sigma_v$ remains roughly constant at values $\sim 2$~km~s$^{-1}$. These results are in fair agreement with the
optical radial velocity survey of \citet{tobin2009}, which covered a
declination range $-6\deg<\delta<-4\deg$.
The velocity dispersion of $^{13}$CO gas is also shown in
Figure~\ref{figure:dec_dispersion}; the stellar $\sigma_v$ is higher than that of the gas the regions of L1641N and around the ONC, whereas between these two, as well as in the southern end of the filament (L1641S), stars show velocity dispersion that are in more agreement with the CO.}

\begin{figure}
\epsscale{1.1}
\plotone{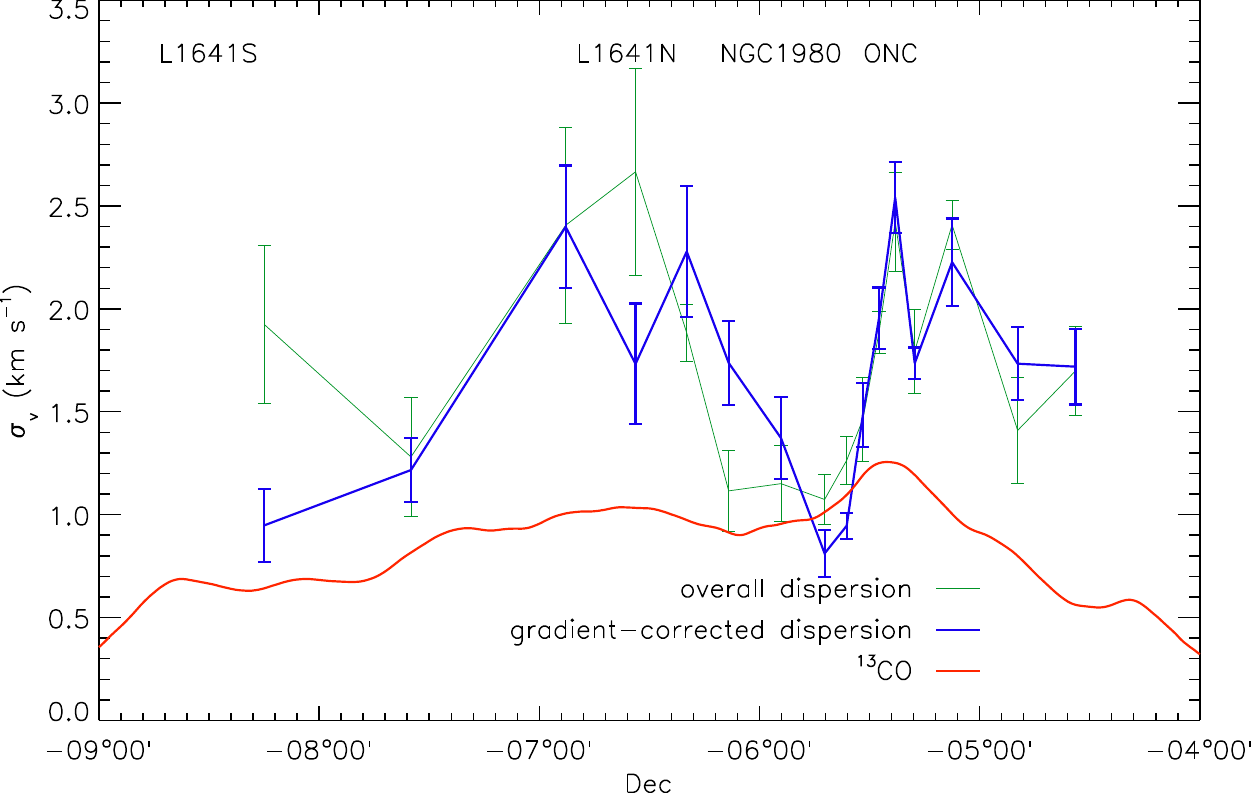}
\caption{
The velocity dispersion $\sigma_v$ as a function of declination. The
blue solid line represents the overall dispersion, corrected for
measurement errors and binarity, in each declination bin. The dashed
blue line is the same with an additional correction for spatial
gradients in $v_r$ in each sample (see text).  The red solid line is
the velocity dispersion of the gas from the survey of
\citet{nishimura2015}. \label{figure:dec_dispersion}}
\end{figure}

\begin{figure}
\epsscale{1.1}
\plotone{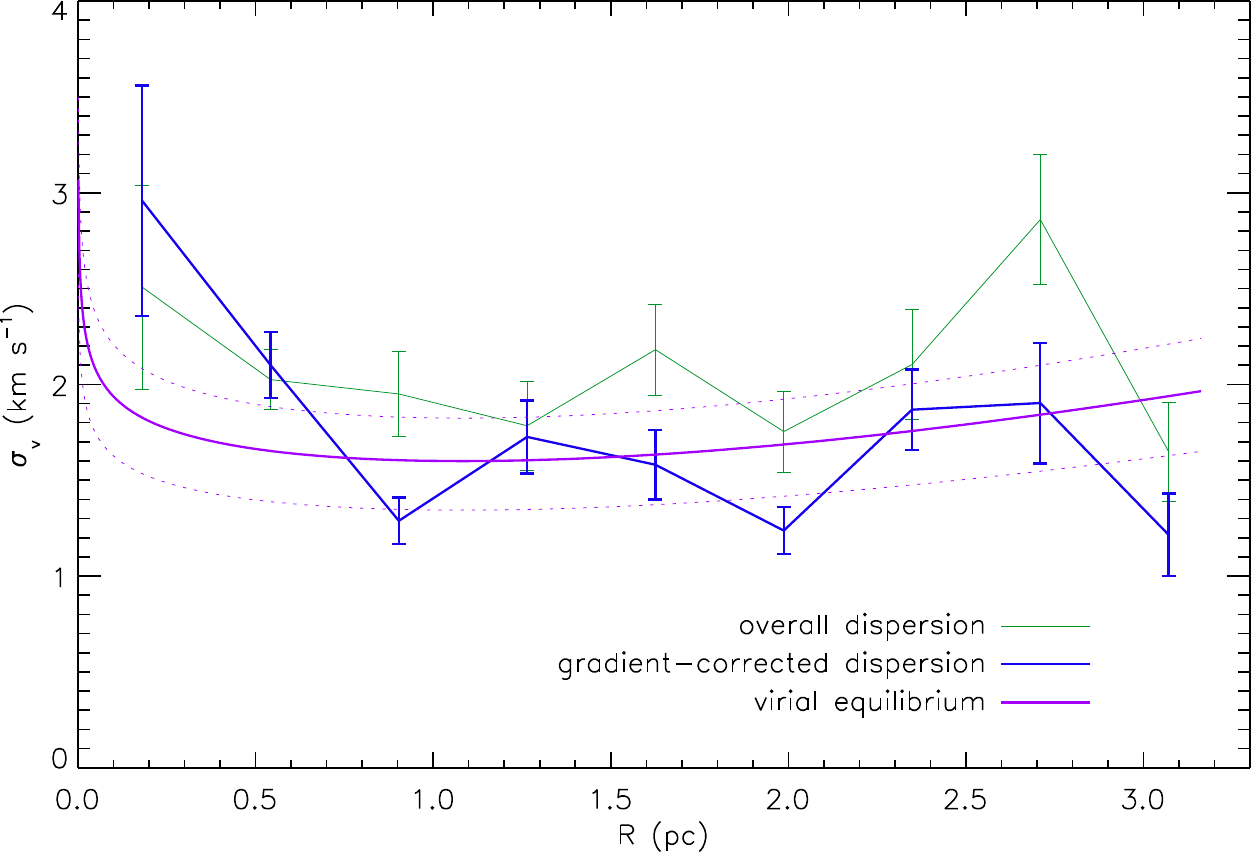}
\caption{
The velocity dispersion $\sigma_v$ for the ONC, measured in radial
annuli at increasing distance from the cluster center. Note that 4~pc
corresponds to $\sim 0.55\deg$ in radius. As in Figure
\ref{figure:dec_dispersion}, the green line is the corrected $\sigma_v$
based on the actual $v_r$, the blue line is computed removing the
large scale spatial gradient in $v_r$. The purple line represents
the equilibrium velocity predicted from the total mass, from
\citet{dario2014b}, and the green dotted line the uncertainty assuming
a 30\% mass uncertainty. \label{figure:ONC_dispersion}}
\end{figure}

Next we focus in more detail on the ONC. Figure
\ref{figure:ONC_dispersion} shows the radial dependence of $\sigma_v$
from the center of the cluster. Again, we compute both the
``standard'' and ``gradient-corrected'' dispersions. { These average at $\sim2.1$ and $\sim1.7$~km~s$^{-1}$ respectively.
Notice, however, that given the small spatial size of the ONC compared to the Orion A cloud, or the width of the declination bins in Figure \ref{figure:dec_dispersion} (3pc correspond to $\sim0.4$\deg), large scale gradients in $v_r$ have a smaller effect on the measured velocity distribution. Still, we do compute the ``gradient-corrected'' $\sigma_v$ along with the ``standard one''.
Our measured standard $\sigma_v$ is in fairly good agreement with the dispersion of $\sim2.3$~km~s$^{-1}$
measured in the ONC by of \citet{tobin2009}, as well as the stellar proper motion
study of \citep{jones-walker1988}; none of these studies, however, had derived $\sigma_v$ after accounting for spatial variations in the mean $\sigma_v$, hence our lower value for the gradient-corrected dispersion.

In Figure \ref{figure:ONC_dispersion} we also show the
predicted velocity dispersion profile required for dynamical virial
equilibrium, as derived in \citet{dario2014b} from the total mass
profile of the ONC. This was obtained combining estimates of the total
stellar mass profile, from censuses of stellar parameters from the
literature, complemented by X-ray and infrared surveys, and estimates
of the gas volume density within the ONC. Specifically, the stellar
component describes a power-law profile slightly steeper than a single
isothermal sphere, and the gas was assumed to be at a constant volume
density of 22~M$_\odot $pc$^{-3}$. This gas density was derived from
the stellar $A_V$ distribution, which averages at $A_V\sim3$ mag, and
not from, e.g., the total column density of molecular material; this
is because the vast majority of the ISM along the line of sight of the
ONC, reaching up to 100~mag \citep{bergin1996,lombardi2014}, lies
behind the cluster. The gas contribution is negligible compared to the
stellar mass up to a few half mass radii ($r_h\simeq 1$~pc,
\citealt{dario2014b}), and is responsible for the increase of the
equilibrium $\sigma_v$ at increasing distances from the center. The
mean $\sigma_v$ in the ONC required for virial equilibrium based on
the total mass is $\sim1.73$~km~s$^{-1}$ \citep{dario2014b}, which,
compared to the measured $2.2$~km~s$^{-1}$
indicates that the ONC is supervirial with a virial parameter \citep{bertoldi1992}
$\alpha_{\rm vir}\simeq1.5$. On the other hand, our gradient corrected $\sigma_v$ indicates a fully virialized system.
The predicted virial velocity is not without uncertainties: in Figure
\ref{figure:ONC_dispersion} we also show the range in the radial
profile of the expected virial velocities assuming an uncertainty of
30\% in the estimated mass. This uncertainty is likely quite large for
the stellar component, being derived from deep samples at different
wavelengths. On the other hand, a higher uncertainty may come from the
gas component, especially if the distribution of the material along
the line of sight---hence the conversion from column densities to
volume density---is quite different than in the orthogonal directions,
specifically, if the system is more flattened along the line of
sight.

Our results show that the ONC population appears in very good agreement with a virial state when considering the gradient-corrected $\sigma_v$, i.e., correcting for local variations in the mean $v_r$.
On the other hand, adopting the standard $\sigma_v$, stars are slightly supervirial. Even in this case, however, velocities are well into the boundness regime ($\sigma_{\rm bound}=\sqrt{2}\sigma_{\rm vir}$).

Even though our results are consisten with a virial state for the ONC, we comment the scenario of a moderately supervirial scenario. This would be
compatible with our presented evidence that the cluster is currently
expanding (see \S\ref{section:expansion}); however a small population of expanding stars can be present even with the bulk of the cluster being virialized.
This could be attribute the some missing mass from recent gas expulsion
from the cluster location. Numerical simulations
\citep[e.g.,][]{farias2015}, however, have shown that the virial state
of a cluster fluctuates significantly during the early evolution of a
stellar system, and the fraction of the stellar population which
remains subsequently bound strongly depends on the dynamical state at
the time of gas removal. Additional data, such as a scheduled
continuation of the IN-SYNC Orion survey, to increase the stellar
sample and obtain additional epochs for our IN-SYNC targets, as well
proper motions from {\it HST} photometry (HST program GO13826, PI
M.~Robberto) or {\it GAIA} for the bright end of the population will
allow us to better characterize this potential expansion of the
ONC. It is also important to stress that a possible infall of material
(gas and stars) along the filament may affect the future fate of the
system. In fact, \citet{tobin2009} suggested that the large velocity
gradient, for both stars and gas, observed to the north of the ONC
(see Figure \ref{figure:radec}) could indicate an infall of material
towards the ONC.

On the other hand, the ONC being close to or fully virialized, is compatible with the evidence that the system is dynamically evolved \citep{dario2014b}.
The fact that $\sigma_v$ for stars in the ONC is larger than the molecular gas dispersion (see Figure \ref{figure:dec_dispersion}) is not of concern. This is because, as anticipated earlier, most of the molecular material in this region of the cloud is located behind the young stellar population, as confirmed by the very low extinction affecting the cluster.
}
We do not find variations of $\sigma_v$ as a function of stellar mass
in the ONC. However, we stress that the IN-SYNC survey does not cover
massive stars, as the sample is limited to below a few $M_\odot$. { We
do measure a slight variation with age: stars younger than the median
age show a $\sigma_v$ which is $0.2$~km~s$^{-1}$ smaller than older stars; this is however significant only at a 1.5$\sigma$ level, and could originate from biases.} On the one hand, older PMS stars
are fainter and thus affected by larger errors; small systematics in
the estimate of the errors in $v_r$ may therefore be responsible for
the detected difference. Also, as shown in \citet{dario2012}, at
fainter luminosities contamination from background sources
increases. Even though we restricted our analysis to known or probable
members, these membership estimates are heterogenous and based on
methods with a range of reliabilities in excluding false candidates.

\begin{figure}
\epsscale{1.1}
\plotone{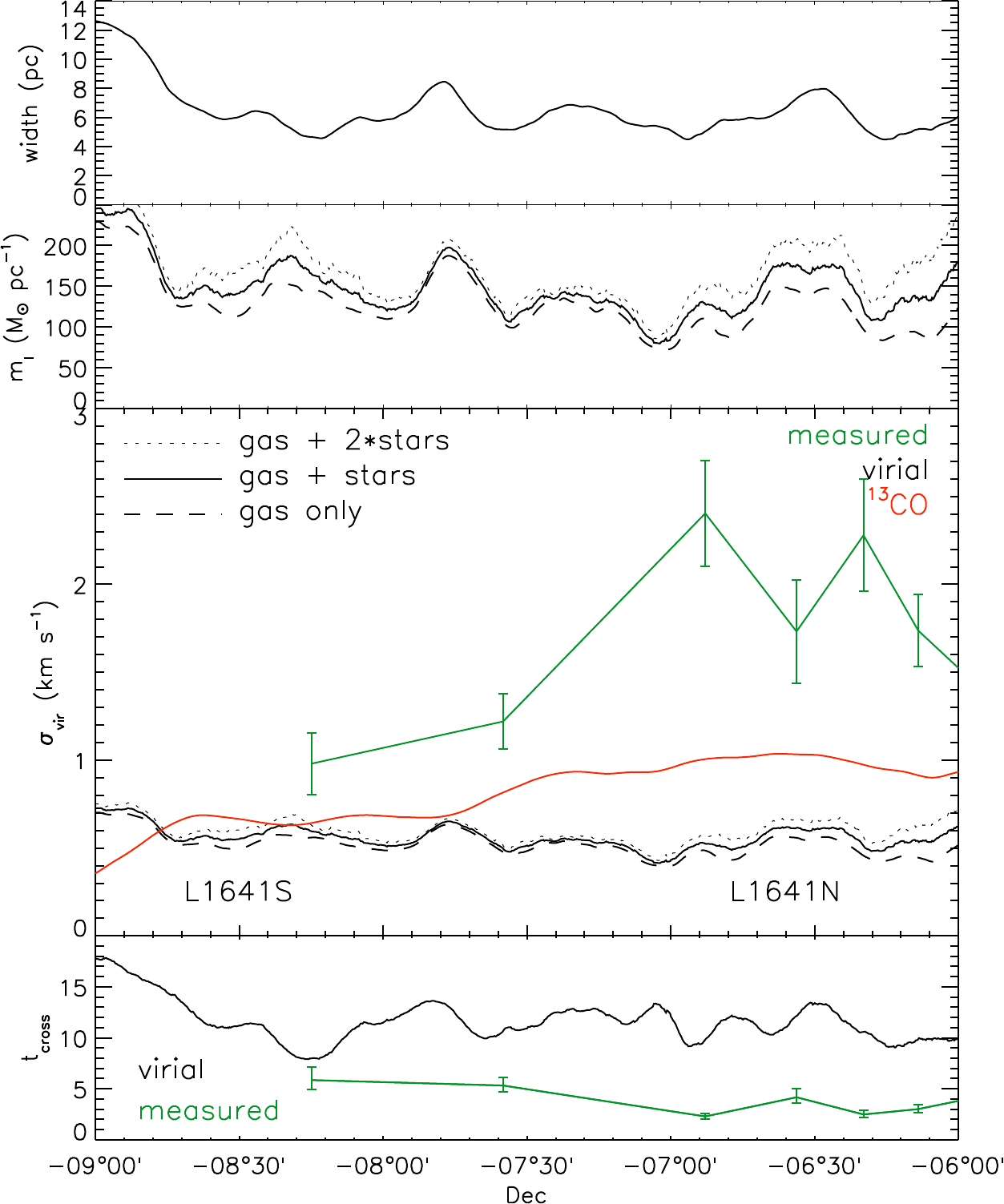}
\caption{
From top to bottom, as a function of declination: a) width of the
filament, defined where $A_V\geq0.3$~$A_V$; b) mass per unit of length
$m_l$; c) virial velocity dispersion $\sigma_{vir}$ from Equation
\ref{equation:filament_virial_sigma}, compared with the observed
dispersions for gas and stars; d) crossing time expected for
$\sigma_{vir}$ compared to the measured
$\sigma_v$. \label{figure:sigma_vir_filament}}
\end{figure}

Returning to the L1641, south of the ONC, Figure
\ref{figure:dec_dispersion} showed that the stellar velocity dispersion in
the northern region (L1641N) is in the range $2$--$2.5$~km~s$^{-1}$, not very dissimilar from
that of the ONC, whereas $\sigma_v$ drops to values $\sim1$~km~s$^{-1}$ in L1641S.
In both these regions, the total density is much lower than
that in the ONC, so that the stellar population should not be bound by
its own mass.
The average stellar density at declinations $\delta<-6\deg$ is
$\sim10$~pc$^{-2}$. Assuming a depth of the population along the line
of sight comparable to the width of the filament in the plane of the
sky ($\sim1.5\deg$ or 10~pc), this results in a stellar number volume
density of $n\lesssim1$~pc$^{-3}$ or
$\rho_*<1\:M_\odot$~pc$^{-3}$. This is more than 2 orders of magnitude
lower than in the ONC, where $\rho_*\simeq250\:M_\odot$~pc$^{-3}$
within $r_h$.


As for the gas density, we adopt the extinction map from
\citet{lombardi2014}; within the Orion A southern filament
($\delta<-6\deg$) the mean extinction is $A_V\simeq 1$~mag. As shown
in \citet{dario2014b}, this corresponds to a mass column density of
$\sim22\:M_\odot$~pc, which again, assuming a depth of the filament of
$\sim10$~pc results in a volume density of molecular material
$\sim2\:M_\odot$~pc$^{-3}$.

We compare these estimates with the condition for virial equilibrium
of a filamentary structure from \citet{fiege2000}, i.e.,
\begin{equation}
\sigma_{vir}\sim\sqrt{\frac{Gm_{l}}{2}}
\label{equation:filament_virial_sigma}
\end{equation}
\noindent where $m_l$ is the mass per unit of length. Considering a
width of the L1641 filament of $\lesssim10$~pc, we obtain a typical
$m_l\sim200\:M_\odot\:$pc$^{-1}$, resulting in $\sigma_{\rm
  vir}\simeq0.65$~km~s$^{-1}$. This is smaller than our measured
$\sigma_v$ throughout the L1641 region. We perform a more rigorous analysis of these quantities in
different positions of the filament. This is shown in Figure
\ref{figure:sigma_vir_filament}, where the displayed quantities, as a
function of declination, have been computed within moving strips of
1~pc in width in the plane of the sky perpendicular to the
filament. The width of the filament has been measured imposing lateral
boundaries where the extinction drops to $A_V=0.3$~mag according to
the map of \citet{lombardi2014}. For the stellar masses at each
position we adopted the values derived in Paper IV; for guidance we
also include the results doubling the stellar masses, as an extreme
case to account for residual incompleteness in our IN-SYNC
sample. This is largely irrelevant, as the total mass is
gas-dominated.

{ The value $\sigma_{\rm vir}\sim0.6$~km~s$^{-1}$ is roughly
constant along the filament; such virial dispersion is several times smaller than the observed $\sigma_v$ in L1641N, indicating that this region is highly unbound.
As shown in Paper IV, here the dust extinction affecting stars is $A_V<1$~mag, and the velocity dispersion highly asymmetric (Section \ref{section:gas-vs-stars}).
Therefore the population here is mostly in foreground of the remaining molecular material, and possibly populated by substructured groups along the line of sight, showing an overall velocity dispersion consistent with being unbound.
On the other hand $\sigma_v$ in L1641S, while being somewhat higher than the virial one, is not incompatible with it, given the uncertainties.

Assuming the present $\sigma_{v}$, the crossing time $t_{\rm cross}$ (see Figure \ref{figure:sigma_vir_filament}) ranges from $\sim3$~Myr in L1641N to $\sim6$~Myr in L1641S. This is somewhat older than than the age of the
system, therefore compatible with the fact that the young
population remains fairly well spatially associated with the molecular
material. As we emphasized above, however, our poor knowledge of the
distribution of material, and in particular molecular gas, along the
line of sight may affect our estimate for the virial equilibrium
velocity predicted for L1641.
}

If the filament is regarded as being a single dynamic entity, then this can explain why the $^{13}$CO velocity dispersion is quite close to the expected virial velocity dispersion. However, then it unclear how the stellar component would have gained such a higher kinetic energy. First, the relatively high dynamical time in the region grants that the stellar population is dynamically young, thus stellar interactions and binary star evolution are nor responsible for the observed velocity dispersion. Also, the L1641 region is deficient of massive stars \citep{hsu2012,hsu2013}, thus a rapid removal of gas from stellar feedback should be ruled out. \citep{stutz2016} recently proposed a model that aims to explain how velocities of young stars can be larger than that of dense gas. They noted that young, but already-formed, stars in Orion A appear more spread out than forming protostars both in the plane of the sky, as well as in their radial velocities (based on IN-SYNC data). The protostars are spatially and kinematically associated to the filaments of molecular gas. They suggested that magnetically induced tangential oscillation of the filament may be taking place. In this scenario, stars are formed coupled with the dense gas, and then gently ejected as the molecular material oscillates back to its original position. However, it is difficult to see how this scenario could lead to observed line of sight velocity dispersions of the $^{13}$CO traced dense gas that are consistently much smaller than those of the ejected stars along the entire filament.

\section{Summary and Conclusions}
\label{section:summary}

We have continued our analysis of the IN-SYNC Orion survey, a
high-resolution near infrared spectroscopic survey of about 2700 stars
in the Orion A star forming region, obtained with the SDSS APOGEE
spectrograph. It covers $\sim6\deg$ in the plane of the sky, or
$\sim40$~pc, thus covering the entire cloud from NGC~1977, the ONC,
L1641N down to L1641S. We cover a mass range from $\sim0.1\:M_\odot$
to a few solar masses and extinctions up to about $20$~mag. In the
previous paper of this series (Paper IV) we presented the derived
stellar parameters, studied the age and age spread of the system, the
reddening law, and assigned new membership estimates for sources
without a previous membership confirmation.

Here we have focussed on the stellar kinematics and dynamics in the
region, from the analysis of radial velocities. Here we summarize the
main results.
\begin{itemize}
\item We find that young stars have average radial velocities that are
  very similar to those of the molecular gas throughout the region as
  traced by CO, following a $\sim10$~km~s$^{-1}$ variation from north
  to south, with the northern region redshifted and the southern
  region blueshifted.
\item In the central part of the cloud, corresponding to L1641N, stars
  appear slightly blueshifted with respect to the gas. In this region,
  however, stars are older and affected by little extinction,
  suggesting that the molecular gas is located behind the population
  along the line of sight, and receding with respect to the stars.
\item In the region of L1641N the distribution of radial velocities
  $v_r$ appears asymmetric, with a broader blue-shifted tail. We
  interpret this as a possible superposition of stellar substructures
  along the line of sight in relative motion with each other.
\item We find the presence of spatial and kinematic substructures in
  the young population throughout the region, as statistically
  significant overdensities in position-position-velocity space. These
  may indicate groupings of stars that have preserved the turbulent
  motions of their natal gas. In the L1641N region, these structures
  show the largest relative offsets in radial velocity, corroborating
  the possibility of a highly structured stellar distribution along
  the line of sight. The ONC, on the other hand, does not present
  significant kinematic substructure, in agreement with previous
  findings for this system to have undergone some degree of dynamical
  evolution, expecting to lead to phase mixing of stellar orbits.
\item We find indications that the ONC may be currently expanding,
  from the correlation between radial velocities and $A_V$, with the
  low-extinction end of the population, closer to the observer along
  the line of sight, blueshifted, and, on the contrary, the high-extinction stars redshifted.
\item { In the ONC, we measure a velocity dispersion
  $\sigma_v\simeq2.2$~km~s$^{-1}$, when considering the measured $v_r$,
  in agreement with previous findings
  in the region based on radial velocities or proper motions.
  On the other hand, when correcting for spatial variations in the mean $v_r$ -- which is critical for large scale gradients in $v_r$, but does marginally affect the results at small scales -- $\sigma_v$ decreases to $\sim1.7$~km~s$^{-1}$. Comparing these values with estimates of the total mass in the
  system, they result in the ONC being, respectively, slightly supervirial (but bound), or in good agreement with a virialized state. Therefore we cannot exclude that the ONC has reached a virial equilibrium.
\item The southern part of the Orion A cloud, L1641, shows two separate behaviors.
The northern part, L1641N, has a velocity dispersion in the range 2--2.5~km~s$^{-1}$, similar to that of the ONC. However, considering the much
  lower density of stars and gas (a few $M_\odot$~pc$^{-3}$) compared
  to the ONC, this is a few times higher than the velocity dispersion
  required for (filamentary) virial equilibrium
  ($\sim0.6$~km~s$^{-1}$), pointing to an unbound population. The low $A_V$ in this region suggests that stars have decoupled from the gas, which is located mostly in the background and had velocities consistent with a virial. In the southern end of the region, L1641S, on the other hand, $\sigma_v$ is lower and in approximate agreement with (or slightly greater than), both a virial state, and the velocity dispersion of the gas. }
\end{itemize}

\acknowledgments
J.J.T acknowledges support from the University of Oklahoma, the Homer L. Dodge
endowed chair, and grant 639.041.439 from the Netherlands
Organisation for Scientific Research (NWO).

\end{document}